\documentclass[conference]{IEEEtran}
\IEEEoverridecommandlockouts
\usepackage{cite}
\usepackage{amsmath,amssymb,amsfonts}
\usepackage{graphicx}
\usepackage{textcomp}
\usepackage{xcolor}
\usepackage{balance}

\usepackage[normalem]{ulem}
\usepackage{soul}
\usepackage{graphicx}
\usepackage{subfig}
\usepackage{comment}
\usepackage{caption}
\usepackage{color}
\usepackage{mathtools}
\usepackage{bm}
\usepackage{array}
\usepackage{lscape}
\usepackage{multirow}
\usepackage{stmaryrd}
\usepackage{url}
\usepackage{boldline}
\usepackage{textcomp}
\usepackage{enumitem}
\usepackage[ruled,vlined,linesnumbered]{algorithm2e}
\usepackage{xcolor}
\usepackage{stmaryrd}
\usepackage{amsmath,amsfonts}
\usepackage[framemethod=tikz]{mdframed}

\DeclareRobustCommand{\IEEEauthorrefmark}[1]{\smash{\textsuperscript{\footnotesize #1}}}

\useunder{\uline}{\ul}{}
\newcommand*{\rred}{\textcolor{black}} 
\newcommand*{\red}{\textcolor{red}}

\definecolor{greencyc}{RGB}{51, 167, 94}
\definecolor{purplecyc}{RGB}{170, 73, 201}
\definecolor{bluecyc}{RGB}{24, 96, 255}
\definecolor{yellowcyc}{RGB}{254, 194, 40}

\newmdenv[innerlinewidth=0.5pt, roundcorner=4pt,linecolor=black,innerleftmargin=6pt,backgroundcolor=gray!20,
innerrightmargin=6pt,innertopmargin=6pt,innerbottommargin=6pt]{graybox}

\newcommand{\model}{TrajCL}

\def\BibTeX{{\rm B\kern-.05em{\sc i\kern-.025em b}\kern-.08em
    T\kern-.1667em\lower.7ex\hbox{E}\kern-.125emX}}

\begin{document}

\title{Contrastive Trajectory Similarity Learning with  Dual-Feature Attention}

\author{
\IEEEauthorblockN{
Yanchuan Chang\IEEEauthorrefmark{1},
Jianzhong Qi\IEEEauthorrefmark{1},\textsuperscript{\textsection}
Yuxuan Liang\IEEEauthorrefmark{2},
Egemen Tanin\IEEEauthorrefmark{1}
}

\IEEEauthorblockA{\IEEEauthorrefmark{1}\textit{The University of Melbourne},
\IEEEauthorrefmark{2}\textit{National University of
Singapore} }
\textit{\{yanchuanc@student., jianzhong.qi@, etanin@\}unimelb.edu.au, 
yuxliang@outlook.com}
}


\maketitle
\begingroup\renewcommand\thefootnote{\textsection}
\footnotetext{Corresponding author}
\endgroup

\begin{abstract}
Trajectory similarity measures act as query predicates in trajectory databases, making them the key player in determining the query results. They also have a heavy impact on the query efficiency. An ideal measure should have the capability to accurately evaluate the similarity between any two trajectories in a very short amount of time.
Towards this aim, we propose a contrastive learning-based trajectory modeling method named \model. 
We present four trajectory augmentation methods and a novel dual-feature self-attention-based trajectory backbone encoder. The resultant model can jointly learn both the spatial and the structural patterns of trajectories. 
Our model does not involve any recurrent structures and thus has a high efficiency. 
Besides, our pre-trained backbone encoder can be fine-tuned towards other computationally expensive measures with minimal supervision data. 
Experimental results show that \model\ is consistently and significantly more accurate than the state-of-the-art trajectory similarity measures. 
After fine-tuning, i.e., to serve as an estimator for heuristic measures, \model\ can even outperform the state-of-the-art supervised method by up to 56\% in the accuracy for processing  trajectory similarity queries. 
\end{abstract}
\begin{IEEEkeywords}
Trajectory similarity, spatial databases, contrastive learning, transformer
\end{IEEEkeywords}

\section{Introduction}\label{sec:intro}
A trajectory is commonly represented as a sequence of location points to describe the movement of an object, such as a person or a vehicle. 
Measuring the similarity between trajectories is a fundamental step in trajectory queries~\cite{DFT,dison,tsjoin,torchdison,torch,stsjoin,dita}, since it is used as a query predicate which determines query results and efficiency.
Unlike numeric data and character data, there are not many universally applicable comparison criteria for trajectory data, and thus measuring similarity between trajectories is an important area of research.

A series of trajectory similarity measures~\cite{edr,edwp,hausdorff,frechet,t2vec,trjsr,cstrm} have been proposed, which can be classified into two categories: \emph{heuristic measures} and  \emph{learned measures}.
Heuristic trajectory similarity measures~\cite{edr,edwp,hausdorff,frechet} mainly aim to find a point-oriented matching between two trajectories based on hand-crafted rules. 
For example, Hausdorff~\cite{hausdorff} leverages the Euclidean distances between points on two trajectories to measure trajectory similarity.
Learned trajectory similarity measures~\cite{t2vec,trjsr,cstrm,e2dtc}, on the other hand, utilize deep learning models to predict similarity values by computing the distance between trajectory-oriented embeddings (i.e., numeric vector representations of trajectories). 
For example, t2vec~\cite{t2vec} and E2DTC~\cite{e2dtc} adapt recurrent neural networks (RNN) to encode trajectories into embeddings, TrjSR~\cite{trjsr} uses convolutional neural networks (CNN) to embed trajectories.

\begin{figure}[ht]
    \subfloat[{Hausdorff (heuristic)}~\label{fig:intro_case1}]{
        \includegraphics[width=0.15\textwidth]{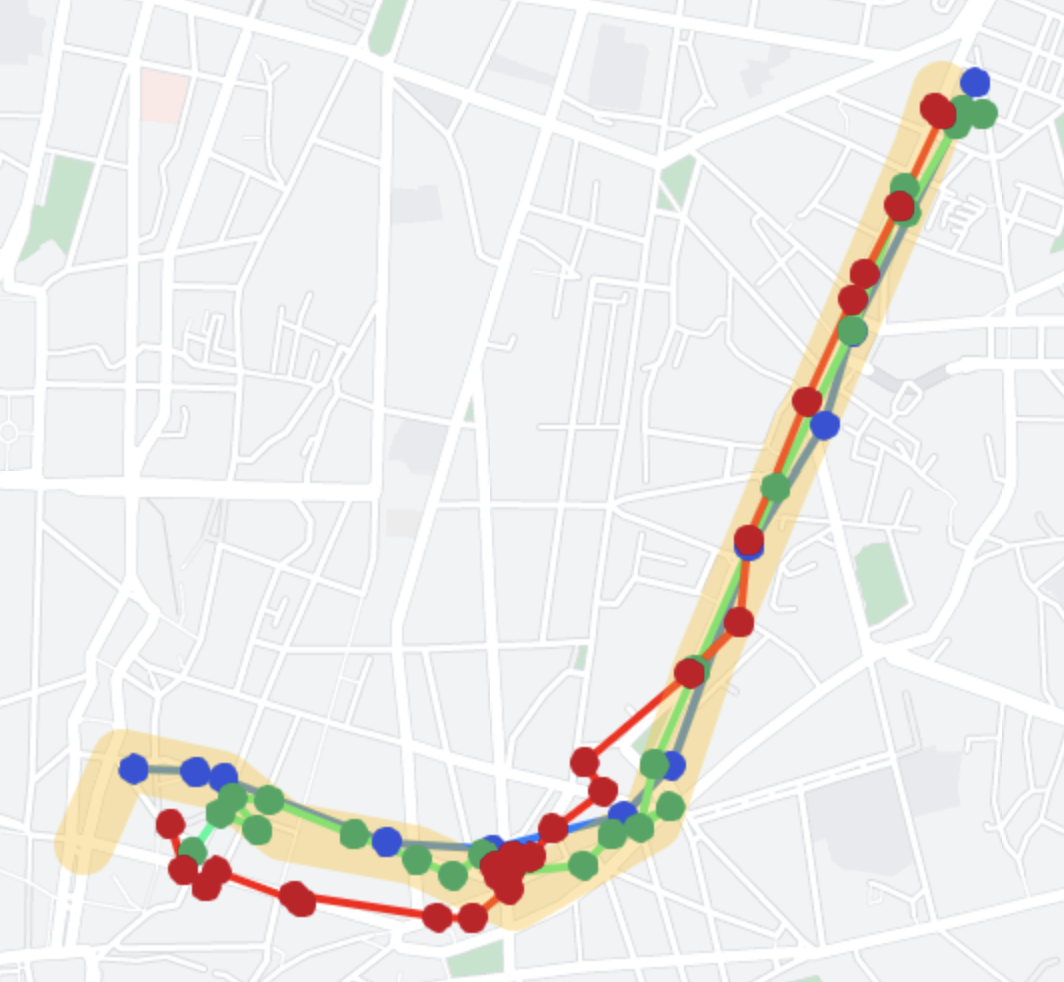}
    } 
    \subfloat[{t2vec (learned)}~\label{fig:intro_case2}]{
        \includegraphics[width=0.15\textwidth]{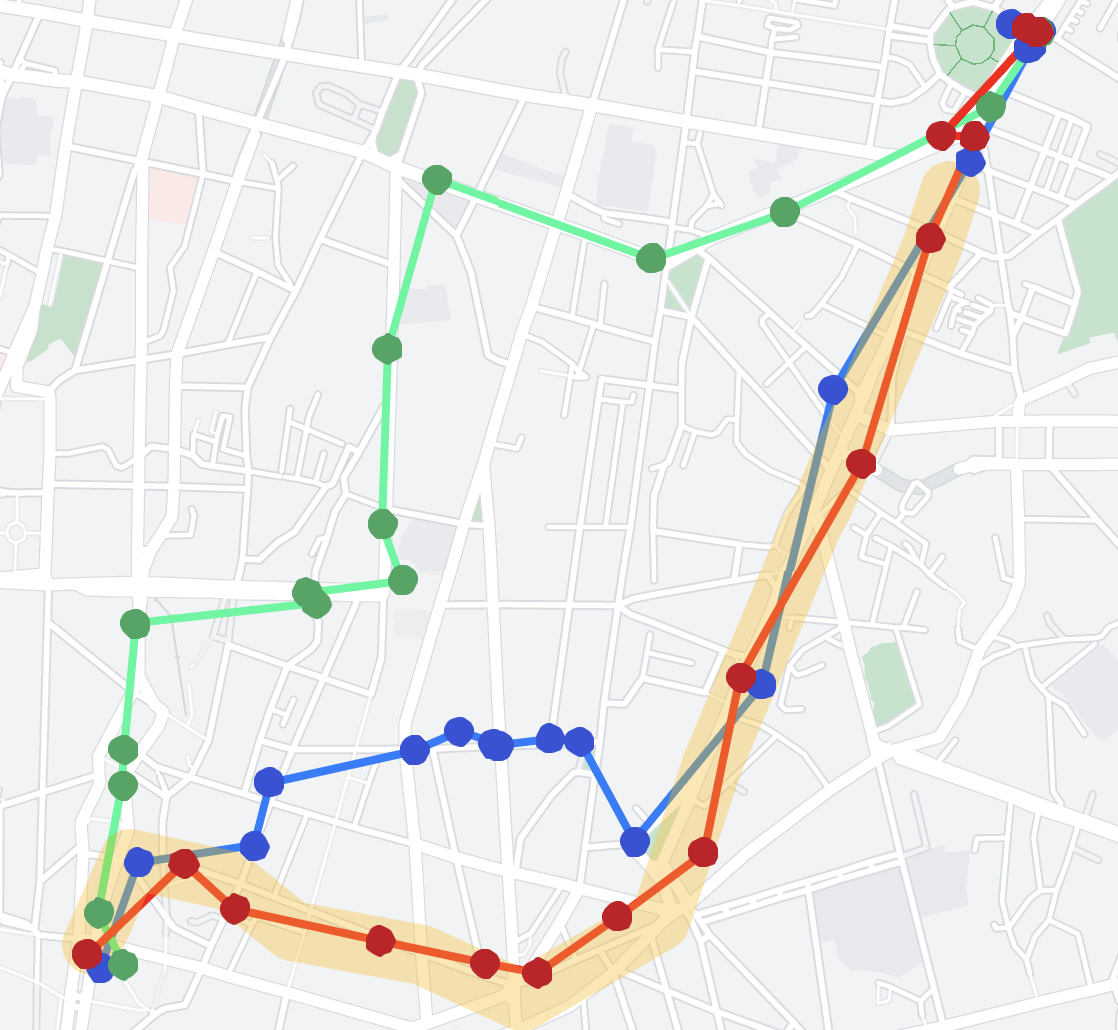}
    } 
    \subfloat[{\model~(ours)}~\label{fig:intro_case3}]{
        \includegraphics[width=0.15\textwidth]{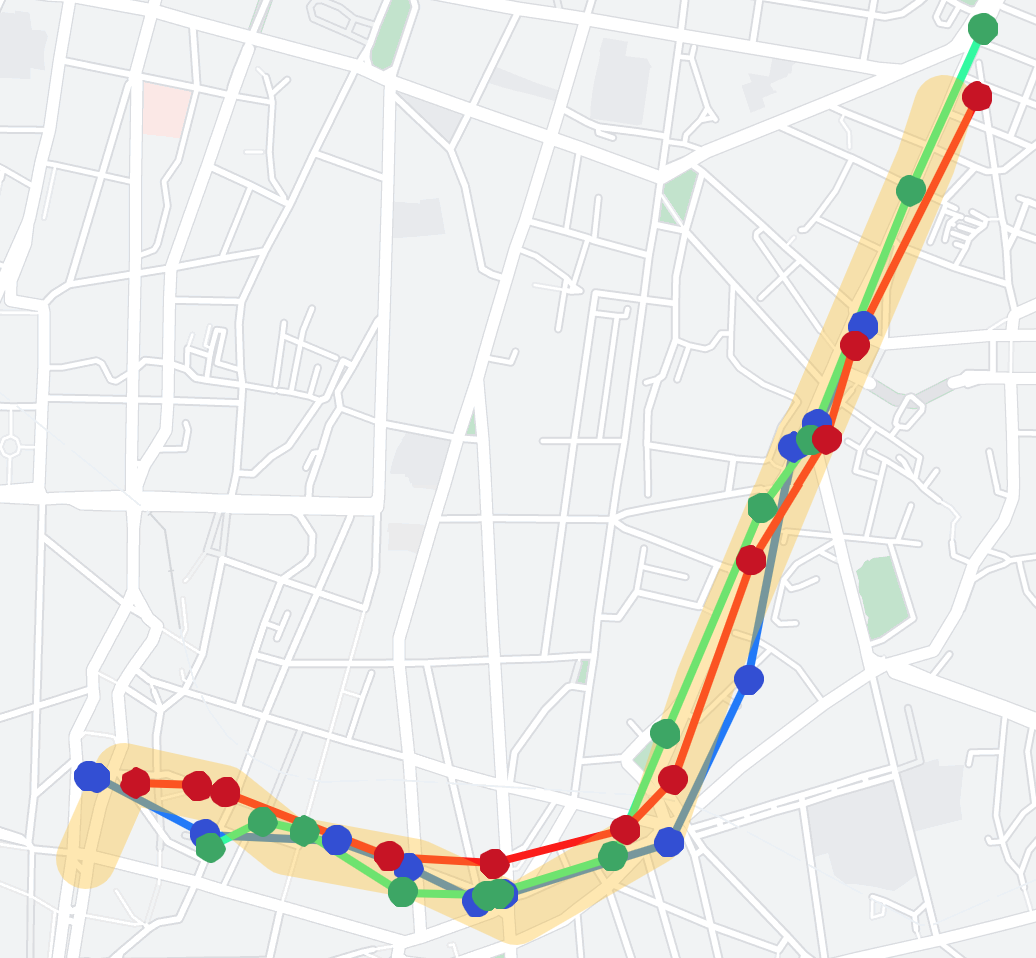}
    }
    \caption{Querying the $3$NN trajectories (The query trajectory is in \textcolor{yellowcyc}{yellow} with extra thick lines for easy viewing. The $3$NN results are colored in \red{red}, \textcolor{greencyc}{green} and \textcolor{bluecyc}{blue}, respectively.) }\label{fig:intro_cases}
\end{figure}

\begin{table}[h]
\centering
\caption{Trajectory similarity computation time}
\label{tab:intro_cases_time}
\begin{tabular}{l|c|c|c}
\hlineB{3}
 & \textbf{Hausdorff} & \textbf{t2vec} & \textbf{TrajCL} \\ \hline \hline
Time ($\mu$s) & 6.63 & 0.34 & 0.14 \\
\hlineB{3}
\end{tabular}
 \vspace{2mm}
\end{table}

Measures in the both categories above face the following challenges.
(1) \textbf{Ineffectiveness}:
Trajectories with different sampling rates or containing noise can degrade the effectiveness of the existing measures.
This is because the heuristic measures using hand-crafted rules are prone to errors by low-quality trajectories. 
The learned measures also suffer from this problem, since they mostly adopt deep learning models which are not originally designed for trajectory data and may fail to capture long spatial correlations between trajectory points and between similar trajectories. 
For example, Fig.~\ref{fig:intro_cases} shows the 3-nearest neighbor query results on the Porto taxi trajectory dataset~\cite{porto}. The query results obtained using t2vec (Fig.~\ref{fig:intro_case2}) are far from the query trajectory. Those obtained by Hausdorff are closer to the query trajectory (Fig.~\ref{fig:intro_case1}), but not as close as those obtained by our \model\ method (Fig.~\ref{fig:intro_case3}), while Hausdorff suffers in efficiency (discussed next).  
(2)~\textbf{Inefficiency}:
Existing heuristic measures compute the distance between each pair of points on two trajectories. They take at least a quadratic time w.r.t. the number of trajectory points, which is unacceptable in online systems, especially when trajectories become longer. 
Although the learned measures get rid of pairwise point comparisons, they are still limited in efficiency.
As Table~\ref{tab:intro_cases_time} shows, Hausdorff takes 6.63 microseconds to compute the similarity of two Porto taxi trajectories. t2vec reduces the time by more than an order of magnitude to  0.34 microseconds. However, its recurrent structure has not fully exploited the parallel power of GPUs. Our \model\ avoids this recurrent structure and further brings the computation time down to 0.14 microseconds.

To address these issues, we propose \model, a \emph{\underline{c}ontrastive \underline{l}earning-based \underline{traj}ectory similarity measure} with a  \emph{dual-feature self-attention-based trajectory backbone encoder} (DualSTB). 
\model\ first leverages our proposed trajectory augmentation methods to generate diverse trajectory variants (i.e., so called views) with different characteristics for each training sample. 
Then, the proposed DualSTB encoder embeds the augmented trajectories into trajectory embeddings, which can capture the spatial distance correlation between the trajectories. After that, we compute the similarity of two trajectories  simply as the $L_1$ distance between their embeddings. 

Due to the lack of ground-truth for trajectory similarity, we train our proposed DualSTB  encoder by adopting self-supervised contrastive learning~\cite{moco,simclr} that aims to maximize the agreement between the representations of positive (i.e., similar) data pairs and minimize that of the negative (i.e., dissimilar) data pairs, where the positive and negative data pairs are generated from input data via augmentation methods.

The idea of using contrastive learning for representation learning is not new. By introducing it into trajectory embedding learning, our first technical contribution is four trajectory augmentation methods that enable obtaining the positive and negative data pairs for 
contrastive learning over trajectories. These methods include point shifting, point masking, trajectory truncating, and trajectory simplification. 
The augmented trajectories can be regarded as a set of low-quality variants of the input trajectories with uncertainty. Such diverse trajectories guide our model to learn the key patterns to differentiate between similar and not-so-similar trajectory pairs.

Our second technical contribution is a \emph{dual-feature self-attention-based trajectory backbone encoder} (i.e., DualSTB) that encodes both structural and spatial trajectory features of a trajectory into its learned embedding. 
The two types of features together provide coarse-grained and fine-grained location information of trajectories.
To obtain a comprehensive embedding based on the two types of features, we devise a dual-feature multi-head self-attention module that 
first learns the correlations between trajectory points based on each type of features. Then, the module adaptively combines the two types of correlations, and finally it forms  the output embeddings.
Such a module can capture the long-term dependency between trajectory points, while its non-recurrent structure enables model inference with high efficiency.

After \model\ is trained, it can be fine-tuned towards any existing heuristic measure as a fast estimator with little training effort, similar to the approximate learned measures~\cite{neutraj,traj2simvec,t3s,trajgat}.
To sum up, we make the following contributions:
\begin{enumerate}[leftmargin=5mm]
    \item We propose \model, a contrastive learning-based trajectory similarity measure that does not rely on any supervision data during training.
    Our measure is robust to low-quality trajectories and efficient on trajectory similarity computation. Besides, pre-trained \model\ models can be used to fast approximate any existing heuristic trajectory similarity measure with little training effort. 
    \item We design four trajectory augmentation methods for our trajectory contrastive learning framework, 
    to enhance the robustness of \model\ on measuring trajectory similarity. 
    \item We present a dual-feature self-attention-based trajectory backbone encoder, which incorporates the structural feature-based attention and the spatial feature-based attention adaptively. It can capture 
    more comprehensive correlations between trajectory points comparing with a vanilla self-attention-based encoder.
    \item We conduct extensive experiments on four trajectory datasets. The results show that: 
    (i) Compared with the state-of-the-art learned trajectory similarity measures, \model\ improves the measuring accuracy by 3.22 times and reduces the running time by more than 50\%, on average.  
    (ii) When acting as a fast estimator of a heuristic measure, \model\ outperforms the state-of-the-art supervised method by up to 56\% in terms of the prediction accuracy.
\end{enumerate}

\section{Related Work}\label{sec:related}

\textbf{Trajectory similarity measures.} 
Existing studies on measuring the similarity between two trajectories can be divided into two categories: heuristic measures and learned measures.

\emph{Heuristic measures}, 
in general, compare pairs of points from two trajectories to find optimal point matches~\cite{euclidean,lcss,edr,erp,edwp,hausdorff,frechet}.
The (Euclidean) distances aggregated from the matched points formulate the similarity of two trajectories. 
Such methods usually take $O(n^2)$ time given trajectories of $n$ points each. 
For example, 
\emph{Hausdorff}~\cite{hausdorff} computes the maximum point-to-trajectory distance between two trajectories.  
\emph{Fr\'echet}~\cite{frechet} resembles Hausdorff but requires the point matches to strictly follow the sequential point order. 
\emph{EDR}~\cite{edr} and \emph{EDwP}~\cite{edwp} compute \emph{edit distance} between trajectories, while \emph{EDwP}~\cite{edwp} further considers the real point distances, and it allows interpolation points to account for non-uniform sampling frequencies.
A few other studies~\cite{dison,tsjoin,torch} measure similarity on spatial networks, which are less relevant and are omitted.

A few recent studies~\cite{neutraj,t3s,traj2simvec,trajgat,gts,st2vec,sarn} take a supervised approach and train a deep learning model to approximate a heuristic measure (e.g., Hausdorff). Once trained, the model can predict trajectory similarity in time linear to the embedding dimensionality. 
For example, \emph{NEUTRAJ}~\cite{neutraj} leverages LSTMs~\cite{lstm} with a spatial memory module to capture the correlation between   trajectories. \emph{Traj2SimVec}~\cite{traj2simvec} accelerates NEUTRAJ training with a  sampling strategy, and it uses an auxiliary loss to capture sub-trajectory similarity. 
\emph{T3S}~\cite{t3s} uses vanilla LSTMs and self-attention~\cite{transformer} to learn heuristic measures.
\emph{TrajGAT}~\cite{trajgat} proposes a graph-based attention model to capture the long-term dependency between trajectories.

\emph{Learned measures}~\cite{t2vec,trjsr,e2dtc,cstrm} \emph{do not require a given heuristic measure to generate model training signals.} 
These methods still learn trajectory embeddings with deep learning, which are expected to 
be more robust to low-quality (e.g., noisy or with low sampling rates) trajectories, since deep learning models are strong in capturing the distinctive data features.  
\emph{t2vec}~\cite{t2vec} uses an RNN-based sequence-to-sequence model to learn trajectory embeddings and then the similarity. It uses a spatial proximity-aware loss that helps encode the spatial distance between trajectories.  
\emph{E2DTC}~\cite{e2dtc} leverages t2vec as the backbone encoder for trajectory clustering. It adds two loss functions to capture the similarity between trajectories from the same cluster.
\emph{TrjSR}~\cite{trjsr} captures the spatial pattern of trajectories by converting trajectories into images.  
\rred{
\emph{CSTRM}~\cite{cstrm} uses vanilla self-attention as its trajectory encoder and proposes a multi-view hinge loss to capture both point-level and trajectory-level similarities between trajectories. It generates positive trajectory pairs using two augmentation methods, i.e., point shifting and point masking, which are empirically shown to be sub-optimal in Section~\ref{sec:exp}.}

Our model is a learned trajectory similarity measure. It aims to address the limitations of the existing learned measures in effectiveness and efficiency as discussed in Section~\ref{sec:intro}.

\textbf{Contrastive learning.}
\emph{Contrastive learning}~\cite{moco,mocov2,simclr,swav,cert,simcse,DeCLUTR,dgi,mvgrl,gcc,gca} is a self-supervised learning technique.  
Its core idea is to maximize the agreement between the learned representations of similar objects (i.e., \emph{positive  pairs})  while minimizing that between dissimilar objects (i.e., \emph{negative  pairs}). The positive and the negative sample pairs are generated from an input dataset, and no supervision (labeled) data is needed.  
Once trained, the representation generation model (i.e., a \emph{backbone encoder}) can be connected to downstream models, to generate object representations  for downstream learning tasks (e.g., classification). 
A few studies introduce contrastive learning into spatial problems, such as traffic flow prediction~\cite{stgcl}.

\textbf{Self-attention models.} 
\emph{Self-attention}-based models~\cite{transformer,bert,ViT,swintransformer} 
learn the correlation between every two elements of an input sequence. 
Studies have adopted self-attention  for trajectory similarity measurement (i.e., T3S and CSTRM).
\rred{Unlike our model, both T3S and CSTRM adopt the vanilla multi-head self-attention encoder~\cite{transformer}, while we propose a dual-feature self-attention-based encoder which can capture trajectory features from two levels of granularity and thus generate more robust embeddings.}

\section{Solution Overview}\label{sec:preliminaries}

We consider a trajectory $T$ as a sequence of points recording discrete locations of the movement of some entity, denoted by $T=[p_1, p_2, ..., p_{|T|}]$, where $p_i$ is the $i$-th point on $T$, and $|T|$ denotes the number of points on $T$. A point $p_i$ is represented by its coordinates in an Euclidean space, i.e., $p_i = (x_i, y_i)$. 

\textbf{Problem statement.} Given a set of trajectories, we aim to learn a trajectory encoder $\mathcal{F}: T \rightarrow \mathbf{h}$ that maps a  trajectory $T$ to a $d$-dimensional embedding vector $\mathbf{h} \in \mathbb{R}^{d}$. The distance between the learned embeddings of two trajectories should be negatively correlated to the similarity between the two trajectories (we use the $L_1$ distance in the experiments). 

\begin{figure*}[t]
    \centering
    \includegraphics[width=0.85\textwidth]{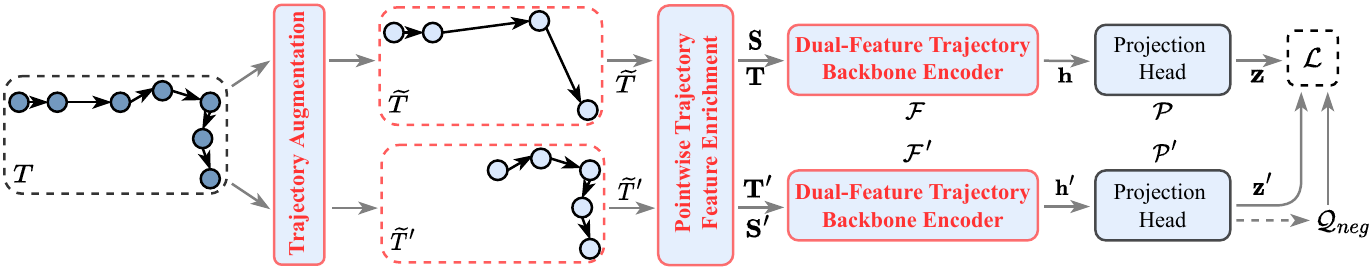}
    \caption{The model architecture of \model\ (The modules in red are our core technical  contributions.)}\label{fig:model_overview}
\end{figure*}

\textbf{Model overview.} 
Fig.~\ref{fig:model_overview} shows an overview of our \model\ model. 
The model follows the dual-branch structure of a strong contrastive learning framework, MoCo~\cite{moco}.  Our technical contributions come in the design of the learning modules as highlighted in red in Fig.~\ref{fig:model_overview}, to be detailed in the next section. 

Given an input trajectory $T$, it first goes through a trajectory augmentation module to generate two different trajectory views (i.e., variants) of $T$, denoted as $\widetilde{T}$ and $\widetilde{T}'$, respectively. We propose four different augmentation methods that emphasize different features of a trajectory (Section~\ref{subsec:augmentation}). The augmentation process is based on $T$ directly, and hence no additional manual data labeling efforts are needed.

The generated views $\widetilde{T}$ and $\widetilde{T}'$ are fed into pointwise trajectory feature enrichment layers to generate pointwise features beyond just the coordinates, which reflect the key characteristics of  $\widetilde{T}$ and $\widetilde{T}'$ (Section~\ref{subsec:dual_input}). We represent the enriched features by two types of embeddings, the \emph{structural feature embedding} and the \emph{spatial feature embedding}, for each point in  $\widetilde{T}$ (and $\widetilde{T}'$). These embeddings encode pointwise structural and spatial features, and form a structural embedding matrix $\mathbf{T}$ ($\mathbf{T'}$) and a spatial embedding matrix $\mathbf{S}$ ($\mathbf{S'}$).

Then, we input ($\mathbf{T}$, $\mathbf{S}$) and ($\mathbf{T'}$, $\mathbf{S'}$) into  \emph{trajectory backbone encoders} $\mathcal{F}$ and $\mathcal{F}'$ to obtain embeddings $\mathbf{h}$ and $\mathbf{h}'$ for $\widetilde{T}$ and $\widetilde{T}'$, respectively (Section~\ref{subsec:backbone}). Our backbone encoders are adapted from
 Transformer~\cite{transformer}, and they encode structural and spatial features of trajectories into the embeddings.

Next, $\mathbf{h}$ and $\mathbf{h}'$ go through two projection heads  $\mathcal{P}$ and $\mathcal{P}'$ (which are fully connected layers of the same structure) to be mapped into  lower-dimensional vectors $\mathbf{z}$ and $\mathbf{z}'$, respectively: 
\begin{equation}\label{eq:projhead}
    \mathbf{z} = \mathcal{P}(\mathbf{h}) = (\mathrm{FC} \circ \mathrm{ReLU} \circ \mathrm{FC})(\mathbf{h})
\end{equation}
Here, $\mathrm{FC}$ denotes a fully connected layer, $\mathrm{ReLU}$ denotes the ReLU activation function, and $\circ$ denotes function composition. We omit the equation for $\mathcal{P}'$ as it is the same. 
Such projections have been shown to improve the embedding quality~\cite{mocov2,simclr}.

\textbf{Model training.} Following previous contrastive learning models, we use the \emph{InfoNCE}~\cite{cpc} loss for model training. We use $\mathbf{z}$ and $\mathbf{z}'$ as a pair of positive samples, as they both come from variants of $T$ and are supposed to be similar in the learned latent space. 
The embeddings (except $\mathbf{z}'$) from projection head $\mathcal{P}'$ that are in the current and recent past training batches are used as negative samples of $\mathbf{z}$. The InfoNCE loss $\mathcal{L}$  maximizes the agreement between positive samples and minimizes that between negative samples: 
\begin{equation}\label{eq:infonce}
\small
\begin{aligned}
    \mathcal{L}(T) = -\log 
    \frac{\mathrm{exp}\big(\mathrm{sim}(\mathbf{z}, \mathbf{z}') / \tau\big)
    }
    {\mathrm{exp}\big(\mathrm{sim}(\mathbf{z}, \mathbf{z}') / \tau\big) +
    \sum_{j = 1}^{|\mathcal{Q}_{neg}|}
    \mathrm{exp}\big(\mathrm{sim}(\mathbf{z}, \mathbf{z}_{j}^-) / \tau\big)
    }
\end{aligned}
\end{equation}
Here, $\mathrm{sim}$ is the cosine similarity. $\tau$ is a \emph{temperature parameter} that controls the contribution of the negative samples~\cite{temperature_parameter}.

We use a queue  $\mathcal{Q}_{neg}$ of a fixed size (an empirical parameter) to store negative samples.
The queue includes the embeddings from $\mathcal{P}'$ in recent batches, to enlarge the negative sample pool, since more negative samples help produce more robust embeddings~\cite{simclr,moco}. 
To reuse negative samples from recent batches, the parameters of $\mathcal{F}'$ and $\mathcal{P}'$ should change smoothly between batches. We follow the \emph{momentum update}~\cite{moco} procedure to satisfy this requirement: 
\begin{equation}\label{eq:momemtum_f}
    \Theta_{\mathcal{F}'} = m\Theta_{\mathcal{F}'} + (1-m)\Theta_{\mathcal{F}}; \ \ \ 
    \Theta_{\mathcal{P}'} = m\Theta_{\mathcal{P}'} + (1-m)\Theta_{\mathcal{P}} 
\end{equation}
Here, $\Theta_{\mathcal{X}}$ denotes the parameters of model $\mathcal{X}$; $m \in (0, 1)$ (which is 0.999 in our experiments) is a  \emph{momentum coefficient} that determines the smoothness of parameter updates. 
Note that $\Theta_{\mathcal{F}}$ and $\Theta_{\mathcal{P}}$ are still updated by stochastic gradient descent.

Once trained, the pointwise trajectory feature enrichment layers and trajectory backbone encoder $\mathcal{F}$ can be detached from  \model\ to serve as an encoder to generate embeddings for given trajectories, which can be used to directly compare the similarity between trajectories. 
They can also be connected to other models  to approximate  heuristic similarity measures.

\section{Model Details}\label{sec:proposed_method} 
We next elaborate our model components,  
including trajectory augmentation methods (Section~\ref{subsec:augmentation}), pointwise trajectory feature enrichment layers (Section~\ref{subsec:dual_input}), and dual-feature  self-attention-based backbone encoders  (Section~\ref{subsec:backbone}).
We also analyze the model complexity (Section~\ref{subsec:complexity}).

\subsection{Trajectory Augmentation}\label{subsec:augmentation}
Data augmentation 
creates different variants of an input record such that the encoder later can learn to capture the common (and distinguishing) features from the variants. 

\begin{figure}[th]
  \centering
  \includegraphics[width=0.44\textwidth]{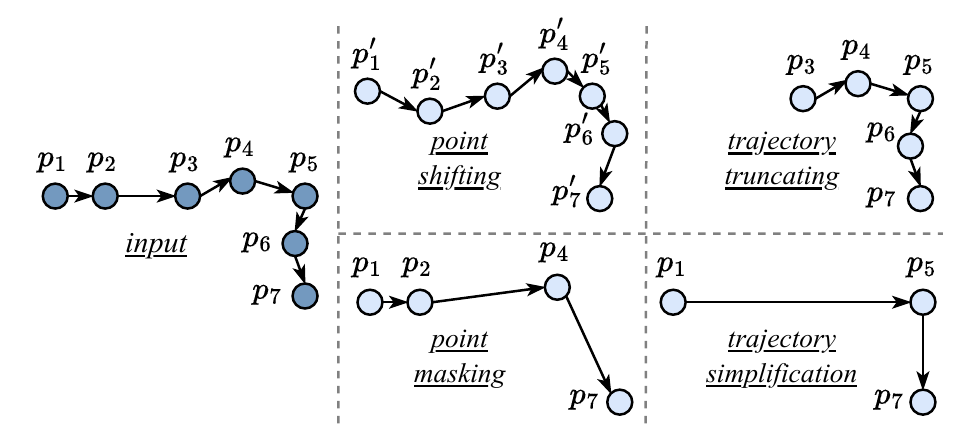}
  \caption{Examples of the proposed trajectory augmentation methods (The same $p_i$ on different trajectories denotes the same point from the input;  $p'_i$ is shifted from $p_i$.)}\label{fig:augmentation}
\end{figure}

No augmentation methods have been proposed for trajectory contrastive learning. We propose four augmentation methods to fill this gap: (1)~\emph{point shifting}, (2)~\emph{point masking} (3)~\emph{trajectory truncating}, and (4)~\emph{trajectory simplification}. 
The aim is to cover the common trajectory transformations. 
Fig.~\ref{fig:augmentation} shows examples for  the four methods, where the trajectory in dark blue denotes an input trajectory, and those in light blue are variants generated by the different augmentation methods.

\textbf{Point shifting.} Given a trajectory $T$, point shifting randomly adds an offset to each coordinate of $p_i \in T$, aiming  to  learn similar trajectories with minor point location differences. The point-shifted output trajectory $\widetilde{T}$ (or $\widetilde{T}'$, same for the rest of the subsection) of $T$ can be represented as: 
\begin{equation}\label{eq:aug_pointshift}
\begin{split}
   \widetilde{T} = [p_1', p_2', \ldots, p_{|T|}'],
   \mathrm{where}\; \forall p_i = (x_i, y_i) \in T,\; \; \; \\
   p_i' = (x_i + \Delta{x_i}, y_i + \Delta{y_i}), \:
   \Delta{x_i} \sim X_n,\: \Delta{y_i} \sim X_n
\end{split}
\end{equation}
Here, $\Delta{x_i}$ and $\Delta{y_i}$ are the location offsets, and $X_n$ is their  distribution. We use a bounded Gaussian distribution $X_n \sim \frac{\rho_m}{\lambda} \cdot \mathcal{N}(\mu, {\sigma}^2)$,  since the location errors of a GPS point cannot be arbitrarily large. 
Parameter $\rho_m$ is the maximum  distance offset, and $\lambda$ is a normalization coefficient to let the integral of the cumulative distribution function of $X_n$ to be 1:
    $\lambda = \int_{-{\rho_m}}^{{\rho_m}} f_{X_n}(x)dx$, 
where $f_{X_n}(x)$ is the probability density function of $X_n$.
We set  $\rho_m$ at 100 meters and use $\mathcal{N}(0, 0.5^2)$ in the  experiments.

\textbf{Point masking.} 
Given a trajectory $T$, point masking randomly masks (i.e., removes) a subset of points in $T$ to generate a variant $\widetilde{T}$, to help learn similar trajectories with varying sampling rates or incomplete records. 
We use an independent and identically uniform distribution for the masking probability of each point, and we set the proportion of points masked, $\rho_d \in (0,1)$, to 0.3 in our experiments. The point-masked output trajectory $\widetilde{T}$ is represented as: 
\begin{equation}\label{eq:aug_pointmask}
\begin{split}
   \widetilde{T} \; = \; [ p_{n_1}, p_{n_2}, \ldots, p_{n_{|\widetilde{T}|}} ],
\end{split}
\end{equation}
where $n_1, n_2, \ldots, n_{|\widetilde{T}|}$ is a strictly increasing sequence,  $\widetilde{T} \subset T$, and  ${|\widetilde{T}|} = \lfloor (1-\rho_{d}) \cdot |T| \rfloor$.

\textbf{Trajectory truncating.} Given a trajectory $T$, trajectory truncating cuts a prefix or a suffix (or both) from $T$ and keeps the rest as a variant $\widetilde{T}$. This method aims to uncover partially overlapped trajectories for applications such as carpooling. We use a parameter $\rho_{b} \in (0, 1)$ to control the proportion of points kept in $\widetilde{T}$. We set $\rho_{b} = 0.7$  in the experiments. Formally, a variant $\widetilde{T}$ generated by trajectory truncating is represented as: 
\begin{equation}\label{eq:aug_subtraj}
\begin{split}
   \widetilde{T} = & [ p_i, p_{i+1}, \ldots, p_{\lfloor i+\rho_{b}\cdot|T| \rfloor}], \; \\ &\text{where } i \text{ is a random integer in } [1, \lceil (1-\rho_b)\cdot |T| \rceil]
\end{split}
\end{equation}

\textbf{Trajectory simplification.}
Given a trajectory $T$, trajectory simplification removes points from $T$ that are not critical to the overall shape and trend of $T$ to form a variant $\widetilde{T}$. The  variant is meant to guide the trajectory encoder to  focus on the critical (e.g., turning) points of $T$. We adopt the \emph{Douglas–Peucker} (DP) simplification algorithm~\cite{rdp} for its wide applicability, although other simplification methods  also apply. DP starts by drawing a line segment to connect the two end points of $T$. The \emph{breaking point} of $T$ that is the farthest from this line segment is calculated (e.g., $p_5$ in Fig.~\ref{fig:augmentation}), and two line segments are drawn to connect this point with the two initial end points, respectively. We repeat the breaking point finding process on each of the two line segments recursively, until the breaking points found are close to the line segments enough (defined by a threshold $\rho_p$ which is 100 meters in the experiments). Only the breaking points found in the process are kept in  $\widetilde{T}$. 
\begin{equation}\label{eq:aug_trajsimp}
   \widetilde{T} = \text{Douglas\_Peucker}(T)
\end{equation}

\textbf{Discussion.} Parameters $\rho_m$, $\rho_d$, $\rho_b$ and $\rho_p$ above control how far off an augmented trajectory can be from the input trajectory. We have set empirical values for them, while changing their values offers flexibility in creating  augmented trajectories that help learn embeddings for trajectory similarity queries of different accuracy requirements.

\subsection{Pointwise Trajectory Feature Enrichment}\label{subsec:dual_input}
After augmenting $T$, we obtain  two augmented trajectory views $\widetilde{T}$ and $\widetilde{T}'$. The next step is to enrich $\widetilde{T}$ and $\widetilde{T}'$ to create features beyond just point coordinates that can reflect the key characteristics of trajectories, which will later be used as the input to the trajectory backbone encoder.

We create two types of features, i.e.,   \emph{structural features} and \emph{spatial features}, for every trajectory point, and we represent each feature by an embedding vector, i.e., the \emph{structural feature embedding} and the \emph{spatial feature embedding}. To also preserve the relative position information of the points, we further encode positional information into embedding vectors.

\textbf{Structural feature embedding.} 
The structural features aim to capture the general shape and point connectivity of a trajectory. We partition the data space with a regular grid where the cell side length is a system parameter, and we represent a trajectory point by the grid cell enclosing it. The sequence of grid cells passed by a trajectory $\widetilde{T}$ (or $\widetilde{T}'$) depicts the trajectory shape, and the cell adjacency relationships reflect the connectivity among the points on the trajectory. 

Using an ID to represent each cell (and the trajectory points inside) offers only sparse information and misses the cell adjacency relationships. Instead, we learn a cell embedding to capture such information as follows. 
We construct a graph where each vertex represents a grid cell. A vertex corresponding to a cell is connected by an edge to each of the eight vertices that correspond to the eight cells surrounding the given cell. We then run a self-supervised graph embedding algorithm (i.e., node2vec~\cite{node2vec}) to learn the vertex embeddings which encode the graph (and hence the grid) structural information. The vertex embeddings are used as the cell embeddings. 

Once the cell embeddings (of $d_t$ dimensions) are obtained, we represent every point $p_i$ on $\widetilde{T}$ (and $\widetilde{T}'$) by the cell embedding of $p_i$. This results in an embedding matrix $\mathbf{T} \in \mathbb{R}^{|\widetilde{T}| \times d_t}$ (and $\mathbf{T'} \in \mathbb{R}^{|\widetilde{T}'| \times d_t}$) to represent $\widetilde{T}$ (and $\widetilde{T}'$).

\textbf{Spatial feature embedding.} 
We further capture fine-grain location information of the points in a trajectory by  computing their spatial feature embeddings.

Given a point $p_i$ on $\widetilde{T}$ (or $\widetilde{T}'$), its spatial feature embedding is a four-tuple $(x_i, y_i, r_i, l_i)$, where $x_i$ and $y_i$ are its spatial coordinates, $r_i$ is the radian between the two trajectory segments before and after $p_i$, i.e., $\overline{p_{i-1}, p_i}$ and $\overline{p_{i}, p_{i+1}}$, respectively, and $l_i$ is the mean length of  $\overline{p_{i-1}, p_i}$ and $\overline{p_{i}, p_{i+1}}$. Formally:
\begin{equation}\label{eq:4tuple}
    r_i  = \angle p_{i-1}p_{i}p_{i+1}; \quad \quad l_i = 0.5 \times\big(|\overline{p_{i-1}, p_i}| + |\overline{p_i, p_{i+1}}| \big)
\end{equation}
We use $\mathbf{S} \in \mathbb{R}^{|\widetilde{T}| \times d_s}$ and $\mathbf{S'} \in \mathbb{R}^{|\widetilde{T}'| \times d_s}$ to denote the spatial feature embedding matrices of $\widetilde{T}$ and $\widetilde{T}'$, respectively ($d_s = 4$).

\textbf{Position encoding.} The structural and the spatial feature embeddings have not considered the relative positions (i.e., preceding and subsequent) of the points on a trajectory, which are important information in trajectory similarity. 
We modify these embeddings to further encode such information. 

We adopt the sine and cosine functions~\cite{transformer} for point-in-sequence position encoding. For the $j$-th dimension value of the structural feature embedding (and the spatial feature embedding) of the $i$-th point on a trajectory, denoted by $\mathbf{T}[i,j]$ (and $\mathbf{S}[i,j]$), we update it by adding the following value $e_{i,j}$: 
\begin{equation}\label{eq:positional}
\begin{split}
    & \mathbf{T}[i,j] = \mathbf{T}[i,j] + e_{i, j}; \quad \quad \mathbf{S}[i,j]  = \mathbf{S}[i,j] + e_{i, j} \\
    e_{i,j} & = \begin{cases}
    \: \sin \big( i / 10000^{j / d_{p}} \big), & j \text{ is even} \\
    \: \cos \big( i / 10000^{(j-1) / d_{p}} \big), & j \text{ is odd}  
    \end{cases}
\end{split}
\end{equation} 
Here, $d_{p}$ denotes the embedding dimensionality of $T$ or $S$, respectively. Intuitively, $e_{i,j}$ encodes the position information $i$ which is added to the embeddings of the $i$-th point.

\subsection{Dual-Feature Self-Attention-Based Backbone Encoder}\label{subsec:backbone}
We propose a \emph{dual-feature self-attention-based trajectory backbone encoder} (DualSTB) equipped with a  \emph{dual-feature multi-head self-attention module} (DualMSM) to capture both structural and  spatial features of an input trajectory. 
Compared with the vanilla multi-head self-attention module (MSM) in Transformer~\cite{transformer}, DualMSM  models not only attentions for each type of features separately but also their joint impact, to generate more comprehensive trajectory representations.

\begin{figure}[th]
  \centering
  \includegraphics[width=0.36\textwidth]{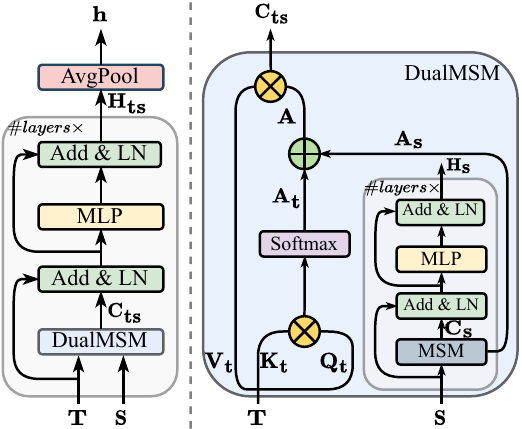}
  \caption{The dual-feature self-attention-based trajectory backbone encoder, DualSTB (left), and the dual-feature multi-head self-attention module, DualMSM (right)}\label{fig:msm}
\end{figure}

\textbf{DualSTB.} Fig.~\ref{fig:msm} shows the structure of DualSTB (the left sub-figure) and its DualMSM module (the right sub-figure). DualSTB  follows the overall structure of a multi-layer Transformer encoder, where our new design lies in DualMSM and the use of two types of input $\mathbf{T}$ and $\mathbf{S}$ instead of one ($\mathbf{T'}$ and $\mathbf{S'}$ for DualSTB  over $\widetilde{T'}$, which is omitted below for conciseness). We include the computation steps of the DualSTB encoder for completeness. Readers who are familiar with the Transformer model can skip the next paragraph. 

DualSTB  takes $\mathbf{T}$ and $\mathbf{S}$ as the input, which go through a DualMSM module to learn the joint attention of the input features. 
The output of DualMSM, denoted by $\mathbf{C_{ts}}$, is fed to a \emph{residual connection}~\cite{residual} (i.e., a \emph{dropout} function~\cite{dropout} and an \emph{add} function) and a \emph{layer normalization}~\cite{layernorm} which help alleviate the problems of gradient vanishing and explosion to obtain smoother gradients. 
Then, the output goes through a \emph{multi-layer perceptron} (MLP) with a residual connection and a layer normalization to slow down the model degeneration. The post-DualMSM processing steps are summarized as follows:
\begin{equation}\label{eq:sublayer1}
    \hat{\mathbf{C}}_\mathbf{ts} = 
    \mathrm{LayerNorm}(\mathbf{T} + \mathrm{Dropout}(\mathbf{C}_\mathbf{ts}))
\end{equation}
\begin{equation}\label{eq:sublayer2}
    \mathbf{H}_\mathbf{ts} = 
    \mathrm{LayerNorm}(\hat{\mathbf{C}}_\mathbf{ts} + \mathrm{Dropout}(\mathrm{MLP}(\hat{\mathbf{C}}_\mathbf{ts})))
\end{equation}
Here, $\hat{\mathbf{C}}_\mathbf{ts}$ is an intermediate result and $\mathbf{H}_\mathbf{ts}$ is the trajectory point representation matrix based on both structural and  spatial features. Multiple layers (two in the experiments) of the structures are stacked. Finally, we apply average pooling  on $\mathbf{H}_{\mathbf{t}\mathbf{s}}$ of the last layer to obtain  trajectory representation $\mathbf{h} \in \mathbb{R}^{d}$,  by  averaging along each dimension of the point representations.

\textbf{DualMSM.} 
The core learning module of the DualSTB encoder, DualMSM, takes as input  both structural features $\mathbf{T} \in \mathbb{R}^{l \times d_t}$ and spatial features $\mathbf{S} \in \mathbb{R}^{l \times d_s}$ of a trajectory and outputs the hidden representations of trajectory points. Here, $l$ denotes the maximum number of points on a trajectory. We pad trajectories with less than $l$ points with 0's.

DualMSM first applies linear transformations on $\mathbf{T}$ to obtain a \emph{value matrix}  $\mathbf{V}_\mathbf{t}^{i} \in \mathbb{R}^{l \times (d_t / h)} = \mathbf{T}\mathbf{W}_{v}^{i}$,
a \emph{key matrix} $\mathbf{K}_\mathbf{t}^{i} \in \mathbb{R}^{l \times (d_t / h)} = \mathbf{T}\mathbf{W}_{k}^{i}$ 
and a \emph{query matrix} $\mathbf{Q}_\mathbf{t}^{i} \in \mathbb{R}^{l \times (d_t / h)} = \mathbf{T}\mathbf{W}_{q}^{i}$, where $h$ is the number of \emph{heads}, $i$ denotes the $i$-th head, and 
$\mathbf{W}_{v}^{i}$, $\mathbf{W}_{k}^{i}$ and $\mathbf{W}_{q}^{i}$ (all in $\mathbb{R}^{d_t \times (d_t / h)}$)
are learnable weights of the $i$-th head. The multi-head mechanism maps different features of $\mathbf{T}$ into different feature sub-spaces such that each head only needs to focus on part of the  features. Each single head attention still covers all features of $\mathbf{T}$, to reduce feature bias.  
Besides, the linear transformation strengthens the representation capacity of the inputs and refrains the attention coefficient matrix from degrading into an identity matrix.

Then, we compute the attention coefficients between the input points of each trajectory: \begin{equation}\label{eq:attention_score}
    \mathbf{A}_\mathbf{t}^{i} = \mathrm{Softmax} \big( \frac{\mathbf{Q}_\mathbf{t}^{i} \mathbf{K}_\mathbf{t}^{{i}\mathsf{T}}}{\sqrt{d_t / h}} \big)
\end{equation}
Here, $\sqrt{d_t / h}$ is used as a scaling factor, and $\mathbf{A}_\mathbf{t}^{i}$ is the \emph{structural attention coefficient matrix}  of the $i$-th head, which represents the structural correlation between the points (cf. the left half of the DualMSM module in Fig.~\ref{fig:msm}). 

For the right half of DualMSM  in Fig.~\ref{fig:msm},  
following a procedure similar to the above, we compute  $\mathbf{V}_\mathbf{s}^{i}$, $\mathbf{K}_\mathbf{s}^{i}$ and $\mathbf{Q}_\mathbf{s}^{i}$ based on the spatial features $\mathbf{S}$ with a different set of learnable parameters, and we compute the  \emph{spatial attention coefficient matrices} $\mathbf{A}_\mathbf{s}^{i}$ following an equation like Equation~\ref{eq:attention_score}. 

Then, we multiply the spatial attention coefficient matrix $\mathbf{A}_\mathbf{s}^{i}$ with the value matrix $\mathbf{V}_\mathbf{s}^{i}$ to obtain the hidden output $\mathbf{C}_\mathbf{s}^{i} \in \mathbb{R}^{l \times (d_s / h)}$ for input $\mathbf{S}$ on the $i$-th head:
\begin{equation}\label{eq:MSA_output}
    \mathbf{C}_\mathbf{s}^{i} = \mathbf{A}_\mathbf{s}^{i} \times \mathbf{V}_\mathbf{s}^{i}
\end{equation} 

After that, we concatenate the output of each head to form the full hidden output $\mathbf{C}_\mathbf{s}$ for $\mathbf{S}$:
\begin{equation}\label{eq:MSA_output_concat}
    \mathbf{C}_\mathbf{s} = 
    (\mathbf{C}_\mathbf{s}^{1} \| \mathbf{C}_\mathbf{s}^{2} \|
    \mathbf{C}_\mathbf{s}^{3} \;... \;\| \mathbf{C}_\mathbf{s}^{h})\mathbf{W}_{o}
\end{equation}
where $\|$ denotes concatenation, and $\mathbf{W}_{o} \in \mathbb{R}^{d_s \times d_s}$ is the learnable weight matrix.  
The steps above on $\mathbf{S}$ correspond to the MSM module at the bottom right of  Fig.~\ref{fig:msm}.

The full hidden output $\mathbf{C}_\mathbf{s}$ then goes through layer normalization with a residual connection, an MLP module with a residual connection, and another layer normalization like those described by Equations~\ref{eq:sublayer1} and~\ref{eq:sublayer2} above, where $\hat{\mathbf{C}}_\mathbf{ts}$, $\mathbf{T}$, $\mathbf{C}_\mathbf{ts}$ and $\mathbf{H}_\mathbf{ts}$ are replaced by $\hat{\mathbf{C}}_\mathbf{s}$, $\mathbf{S}$, $\mathbf{C}_\mathbf{s}$ and $\mathbf{H}_\mathbf{s}$, respectively.  
Similar to Transformer, we stack these layers in DualMSM (two layers in the experiments).

So far, we have obtained both the structural and the spatial  attention coefficient matrices  $\mathbf{A}_\mathbf{t}^{i}$ and   $\mathbf{A}_\mathbf{s}^{i}$ ($\mathbf{A}_\mathbf{s}^{i}$ denotes the matrix of the last stacked layer). 
We calculate a weighted sum of  $\mathbf{A}_\mathbf{t}^{i}$ and $\mathbf{A}_\mathbf{s}^{i}$ with a learnable weighting parameter $\gamma$ as the final attention coefficient. This learnable parameter guides the learned embeddings to adaptively take structural and spatial correlations between trajectory points into consideration. 
The output of the $i$-th head of DualMSM, denoted as   $\mathbf{C}_{\mathbf{t}\mathbf{s}}^{i} \in \mathbb{R}^{l \times (d_t / h)}$, is computed by:
\begin{equation}\label{eq:dualMSA_output}
    \mathbf{C}_{\mathbf{t}\mathbf{s}}^{i} = \big( \mathbf{A}_\mathbf{t}^{i} + \gamma \mathbf{A}_\mathbf{s}^{i})\mathbf{V}_\mathbf{t}^{i}
\end{equation}

Lastly, similar to $\mathbf{C}_{\mathbf{s}}^{i}$ in Equation~\ref{eq:MSA_output_concat}, we concatenate the output of each head $\mathbf{C}_{\mathbf{t}\mathbf{s}}^{i}$ and apply a  linear transformation to form the final output of DualMSM, i.e., $\mathbf{C}_{\mathbf{t}\mathbf{s}} \in \mathbb{R}^{l \times d_t}$.

\textbf{Discussion.}
The differences between DualMSM  and the vanilla MSM~\cite{transformer} are twofold.
First, DualMSM takes as input two types of  features, i.e., structural and spatial, for trajectory embedding learning, while MSM only accepts one type of input features. One may concatenate both types of features into one to suit the input structure of MSM. Such an approach, however, is shown to be inferior empirically  (Section~\ref{subsec:ablation}).
Second, DualMSM allows learning exclusive attention coefficients for each type of input features, which are then integrated 
adaptively. Such a mechanism is not supported in MSM. This mechanism ensures that the correlations (i.e., the attention coefficients) between points based on different types of features are modeled independently, while the adaptive integration makes the attention mechanism more flexible to combine different types of input features.

\subsection{Cost Analysis}\label{subsec:complexity}
\model\ takes $O(l^2 \cdot d \cdot L)$ time to compute $\mathbf{h}$ for $T$,
where $l$ denotes the number of points on $T$, and $L$ is the number of DualMSM layers.  
This cost has hidden the $O(l)$ time to obtain the pointwise trajectory features, i.e., $T \rightarrow \mathbf{T}$ and $\mathbf{S}$, since the dominating cost comes from the trajectory encoding stage (i.e., $\mathbf{T}$ and $\mathbf{S} \rightarrow \mathbf{h}$), in particular, the matrix multiplication costs in Equations~\ref{eq:attention_score} and~\ref{eq:MSA_output}.
Without any recurrent structures, our \model\ model can be easily  accelerated by GPUs. 

In comparison, the representative competitor methods t2vec~\cite{t2vec} and E2DTC~\cite{e2dtc} take $O(l \cdot d^2 \cdot L)$ time to compute a trajectory embedding, TrjSR~\cite{trjsr} takes $O(m^2 \cdot k^2 \cdot n_k \cdot cl \cdot L)$ time, and CSTRM~\cite{cstrm} takes the same time as \model, where $m$ is the side length (number of pixels) of trajectory images, $k$ is the side length of convolution kernels, $n_k$ is the number of kernels, and $cl$ is the number of image channels in TrjSR. 

Once $\mathbf{h}$ is obtained, it will
only take a time linear to $d$
to compute the similarity between two trajectories.

\section{Experiments}\label{sec:exp}
We evaluate \model\ on four real trajectory datasets by comparing with heuristic methods and learned methods for both trajectory similarity computation and similarity queries.
We also study the effectiveness of \model\ on a downstream task by fine-tuning it to learn  heuristic similarity measures. 
Finally, we study the impact of model components and parameters.

\subsection{Experimental Settings}\label{subset:exp_settings}
\textbf{Datasets.} 
We use four real-world trajectory datasets: (1)~\textbf{Porto}~\cite{porto} contains 1.7 million taxi trajectories from Porto, Portugal, between July 2013 and June 2014; (2)~\textbf{Chengdu}~\cite{didi},  contains 6.1 million ride-hailing trajectories from Chengdu, China, during November 2016; (3)~\textbf{Xi'an}~\cite{didi} contains 2.1 million ride-hailing trajectories from Xi'an, China, during the first two weeks of October 2018; and
\rred{(4)~\textbf{Germany}~\cite{osm_route} is a country-wide dataset, containing 170.7 thousand user-submitted trajectories within Germany, between 2006 and 2013.}
Following previous studies~\cite{t2vec,trjsr,cstrm}, we preprocess each dataset by filtering out  trajectories that are outside the city area or contain less than 20 points or more than 200 points. 
The datasets after preprocessing are summarized in Table~\ref{tab:dataset_statistics}.

\begin{table}[h]
\centering
\caption{Dataset statistics}
\label{tab:dataset_statistics}
\resizebox{\columnwidth}{!}{%
\begin{tabular}{l|r|r|r|r}
\hlineB{3}
 & \textbf{Porto} & \textbf{Chengdu} & \textbf{Xi'an}  & \textbf{Germany} \\ \hline \hline
\#trajectories & 1,372,725 & 4,477,858 & 900,562 & 143,417\\
Avg. \#points per trajectory & 48 & 105 & 118 & 72\\
Max. \#points per trajectory & 200 & 200 & 200 & 200\\
Avg. trajectory length (km) & 6.37 & 3.47 & 3.25 & 252.49 \\
Max. trajectory length (km) & 80.61 & 23.12 & 99.41 & 115,740.67 \\ 
\hlineB{3}
\end{tabular}
} 
\end{table}

\rred{Each dataset is randomly partitioned into four disjoint subsets: (1)~200,000 trajectories in Porto, Chengdu and Xi'an, and 30,000 trajectories in Germany for training, respectively, (2)~a 10\% subset for validation, (3)~100,000 trajectories for testing, and (4)~10,000 trajectories for downstream task experiments, i.e., learning to approximate a heuristic similarity measure, which are further split by 7:1:2 for training, validation, and testing.}

\textbf{Competitors.} We compare our method \model\ with four representative heuristic trajectory similarity measures, i.e., \textbf{EDR}~\cite{edr}, \textbf{EDwP}~\cite{edwp}, \textbf{Hausdorff}~\cite{hausdorff} and \textbf{Fr\'echet}~\cite{frechet}, and three recently proposed self-supervised learned measures, i.e., \textbf{t2vec}~\cite{t2vec}, \textbf{TrjSR}~\cite{trjsr}, \textbf{E2DTC}~\cite{e2dtc} and \textbf{CSTRM}~\cite{cstrm}
Similar to \model, all these measures are used as a standalone trajectory similarity measure. These  baseline methods are described in  Section~\ref{sec:related}.

In the downstream task, i.e., fine-tuning \model\ to learn and approximate a heuristic measure, we compare \model\ with the above self-supervised learned methods, as well as the latest supervised methods \textbf{Traj2SimVec}~\cite{traj2simvec}, \textbf{T3S}~\cite{t3s} and \textbf{TrajGAT}~\cite{trajgat}. 
We omit NEUTRAJ~\cite{neutraj} as it has been shown to be outperformed by these methods already.

We use the released code and default parameters for all baseline methods except Traj2SimVec and T3S which have no released code. We implement these two methods following their original proposals~\cite{traj2simvec,t3s}.

\textbf{Implementation details.} We implement \model\footnote{Code is available at \url{https://github.com/changyanchuan/TrajCL}} with PyTorch 1.8.1. We run experiments on a machine with a 32-core Intel Xeon CPU, an NVIDIA Tesla V100 GPU and 64 GB RAM. We report mean results over five runs of each experiment with different random seeds. 

We train \model\ using the Adam optimizer. The maximum number of training epochs is set to 20, and we early stop after 5 consecutive epochs without improvements in the loss. The learning rate is initialized to 0.001 and decayed by half after every 5 epochs. We set the embedding dimensionality $d$ to 256 for all learned methods except TrajGAT, which performs better with its default $d$ value 32 and hence has used this value.  
For \model\ and T3S, the number of heads $h$ is 4, and the number of encoder layers $\#layers$ is 2.
The default augmentation methods are point masking and trajectory truncating for the two views. The side length of the grid cells is 100 meters.

We use `$\blacktriangle$' (and `$\blacktriangledown$') to indicate that larger (and smaller)  values are better, and the best results are in bold.

\subsection{Learning Trajectory Similarity}\label{subsec:exp_newsimi}
We first investigate the effectiveness of \model\ on learning trajectory similarity, i.e., to find similar trajectories.

\textbf{Setup.} 
\rred{Following the baseline methods~\cite{t2vec,trjsr,cstrm}}, the experimental data includes a \emph{query set} $Q$ and a \emph{database} $D$, which are created from the 100,000 randomly chosen testing set (see Datasets in Section~\ref{subset:exp_settings} above) as follows. We test how well \model\ can help recover the ground-truth similar trajectories in $D$ for the query trajectories in $Q$. 

From each dataset, we randomly sample 1,000 trajectories from the 100,000 testing set. For each sampled trajectory $T^q$, we create two sub-trajectories -- one consisting of the odd points of $T^q$, i.e.,  $T^q_a=[p_1, p_3, p_5, \ldots]$, and the other the even points, i.e.,  $T^q_b=[p_2, p_4, p_6, \ldots]$. Trajectory $T^q_a$ is put into the query set $Q$, and $T^q_b$ is put into the database $D$ and will serve as the ground-truth most similar trajectory of $T^q_a$. We further add randomly chosen trajectories from the testing set into $D$ to form databases of different sizes. 
\rred{We generate $T^q_a$ and $T^q_b$ because there is no known ground-truth similar trajectory pair. Such a pair can be seen as different trajectories recorded with the same sampling rate for the same movement sequence but starting at slightly different locations. Thus, they can be considered as a reasonably similar pair.}

For every query $T^q_a \in Q$, we compute the similarity  between $T^q_a$ and all trajectories in $D$ (for each method),  and we report the \textbf{mean rank} of $T^q_b$ by sorting the similarity values of trajectory pairs in descending order. Ideally, $T^q_b$ should rank 1$^{st}$, for the reason above.

\begin{table}[ht]
\centering
\caption{Mean rank ($\blacktriangledown$) of the ground truth most similar  trajectory vs. database size (Best results are in \textbf{bold}.)}
\label{tab:newsimi_dbsize}
\resizebox{\columnwidth}{!}{%
\begin{tabular}{l|l|rrrrr}
\hlineB{3}
\textbf{Dataset} & \textbf{Method} & \textbf{20K} & \textbf{40K} & \textbf{60K} & \textbf{80K} & \textbf{100K} \\ \hline \hline
\multirow{9}{*}{\textbf{Porto}} & EDR & 8.318 & 14.398 & 17.983 & 22.902 & 28.753 \\
 & EDwP & 3.280 & 4.579 & 5.276 & 6.191 & 7.346 \\
 & Hausdorff & 3.068 & 4.014 & 4.649 & 5.451 & 6.376 \\
 & Fr\'echet & 3.560 & 4.959 & 5.968 & 7.192 & 8.631 \\ \cline{2-7} 
 & t2vec & 1.523 & 2.051 & 2.257 & 2.612 & 3.068 \\
 & TrjSR & 1.876 & 2.783 & 3.208 & 3.826 & 4.635 \\
 & E2DTC &  1.560 &	2.111 &	2.349 &	2.731 &	3.213 \\
 & \rred{CSTRM} &  \rred{4.476} & \rred{7.954} & \rred{10.630} & \rred{13.576} & \rred{16.699} \\
 & TrajCL & \textbf{1.005} & \textbf{1.006} & \textbf{1.006} & \textbf{1.007} & \textbf{1.010} \\ \hline \hline
\multirow{9}{*}{\textbf{Chengdu}} & EDR & \:\:\, 80.021 & 156.847 & 234.128 & 312.528 & 388.796 \\
 & EDwP & 9.066 & 15.890 & 22.779 & 29.615 & 36.724 \\
 & Hausdorff & 25.248 & 48.293 & 71.537 & 94.861 & 118.179 \\
 & Fr\'echet & 18.973 & 35.731 & 52.622 & 69.617 & 86.620 \\ \cline{2-7} 
 & t2vec & 4.705 & 8.225 & 11.811 & 15.461 & 19.114 \\
 & TrjSR &  4.421 & 7.723 & 11.027 & 14.364 & 17.873  \\
 & E2DTC &  5.062 & 8.880 & 12.816 & 16.784 & 20.775\\
 & \rred{CSTRM} & \rred{5.914} & \rred{10.655} & \rred{15.423} & \rred{20.100} & \rred{24.900} \\
 & TrajCL & \textbf{1.038} & \textbf{1.074} & \textbf{1.113} & \textbf{1.151} & \textbf{1.200} \\ \hline \hline
\multirow{9}{*}{\textbf{Xi'an}} & EDR & 57.149 & 113.583 & 169.284 & 224.900 & 280.126 \\
 & EDwP & 2.318 & 2.611 & 2.929 & 3.288 & 3.606 \\
 & Hausdorff & 37.896 & 74.044 & 109.996 & 145.924 & 182.224 \\
 & Fr\'echet & 40.378 & 79.087 & 117.677 & 156.159 & 194.685 \\ \cline{2-7} 
 & t2vec & 2.574 & 4.047 & 5.538 & 7.047 & 8.644 \\
 & TrjSR & 13.791 & 26.901 & 39.683 & 52.559 & 65.647 \\
 & E2DTC & 2.988 & 4.909 & 6.854 & 8.810 & 10.861 \\
 & \rred{CSTRM} & \rred{3.078} & \rred{5.231} & \rred{7.317} & \rred{9.402} & \rred{11.635} \\
& TrajCL & \textbf{1.023} & \textbf{1.050} & \textbf{1.066} & \textbf{1.087} & \textbf{1.107} \\ \hline \hline
 
\multirow{9}{*}{\textbf{\rred{Germany}}} &  \rred{EDR} & \:\:\, \rred{279.385} & \rred{558.288} & \rred{834.208} & \rred{1108.975} & \rred{1370.004} \\
 & \rred{EDwP} & \rred{2.168} & \rred{2.277} & \rred{2.371} & \rred{2.454} & \rred{2.515} \\
 & \rred{Hausdorff} & \rred{2.803} & \rred{3.509} & \rred{4.206} & \rred{4.906} & \rred{5.551} \\
 & \rred{Fr\'echet} & \rred{2.581} & \rred{3.108} & \rred{3.633} & \rred{4.113} & \rred{4.589} \\ \cline{2-7} 
 & \rred{t2vec} & \rred{1.571} & \rred{1.982} & \rred{2.387} & \rred{2.718} & \rred{3.053} \\
 & \rred{TrjSR} &  \rred{6.517} & \rred{11.741} & \rred{16.969} & \rred{22.182} & \rred{24.083}  \\
 & \rred{E2DTC} &  \rred{3.136} & \rred{5.156} & \rred{7.248} & \rred{9.207} & \rred{10.956} \\
 & \rred{CSTRM} &  \rred{-} & \rred{-} & \rred{-} & \rred{-} & \rred{-} \\
 & \rred{TrajCL} & \rred{\textbf{1.012}} & \rred{\textbf{1.022}} & \rred{\textbf{1.034}} & \rred{\textbf{1.040}} & \rred{\textbf{1.045}} \\ 
\hlineB{3}
\end{tabular}
}
\end{table}

\textbf{Results.} \emph{Varying database size $|D|$.}
We first vary the database size $|D|$ from 20,000 to 100,000.  Table~\ref{tab:newsimi_dbsize} shows the mean rank of $T^q_b$ (smaller values are better) produced by the different methods. 
\model\ outperforms all heuristic and learned similarity measures on all four datasets, producing mean ranks very close to 1 consistently. 
For example, on Porto, compared with those of the best heuristic competitor Hausdorff and the best learned competitor t2vec, the mean rank of $T^q_b$ for \model\ is up to 5.31 times and 2.04 times smaller (1.010 vs. 6.376 and 3.068 when $|D|$=100K), respectively. 
Further, \model\ shows better stability with respect to $|D|$. When $|D|$ grows,  \model\  can still tell that $T^q_b$ is more similar to $T^q_a$ than any other trajectories added to $D$, with the worst mean rank of $T^q_b$ being 1.200, which is captured on Chengdu when $|D|$=100K. In comparison, the worst-case mean rank of $T^q_b$ of the best baseline method (i.e., TrjSR) on the same dataset grows to 17.873 which is 13.89 times larger. Such results confirm the effectiveness of \model\ to obtain better trajectory representations that  preserve the similarity.

We note that t2vec and E2DTC share similar results, as they use the same backbone encoder. E2DTC is slightly worse than t2vec even though it is a newer method. This is because E2DTC is designed for trajectory clustering which may not be optimized for trajectory similarity learning. We also note that TrjSR has reported better performance than t2vec~\cite{trjsr}. However, we were not able to produce the same results on our datasets, while we do not have access to their datasets.

\rred{Besides, although CSTRM also uses self-attention, it cannot accurately learn trajectory similarity. This is because CSTRM uses the vanilla MSM and only learns coarse-grained trajectory representations based on grid cells, while our proposed DualMSM can capture both coarse-grained and fine-grained features and leverage the topology of grid cells. Further, due to the large number of parameters of the cell embedding module in CSTRM, it triggers an out-of-memory error on Germany and hence no results were  obtained.}

\rred{Further, we observe that, in general, the learning-based methods achieve better results on Germany than on Chengdu and Xi'an, and best results on Porto. Germany has the largest geographical region and grid space among the four datasets, while it has the fewest training trajectories. These make trajectory point correlation learning among different grid cells difficult and lead to a more challenging dataset than Porto. On the other hand, Germany has the lowest trajectory density, and its trajectories are easier to be distinguished among each other, especially comparing with those in the Chengdu and Xi'an datasets which are much denser (with the smaller spatial regions).}

\emph{Varying down-sampling rate $\rho_{s}$.} 
We down-sample trajectories in $Q$ and $D$ by randomly masking points in each trajectory with a probability $\rho_{s} \in [0.1, 0.5]$, while 
$|D| = 100,000$.
Table~\ref{tab:newsimi_downsampling} shows the performance results.  
\model\ again achieves the smallest mean ranks of $T^q_b$ consistently. 
Compared with the best heuristic competitor EDwP and the best learned competitor t2vec, \model\ reduces the mean rank of $T^q_b$ by at least 0.87 times and 2.17 times (on Porto), while the advantage is up to 11.38 times and 12.01 times, respectively.

\begin{table}[ht]
\centering
\caption{Mean rank ($\blacktriangledown$) vs. down-sampling rate}
\label{tab:newsimi_downsampling}
\resizebox{\columnwidth}{!}{%
\begin{tabular}{l|l|rrrrr}
\hlineB{3}
\textbf{Dataset} & \textbf{Method} & \textbf{0.1} & \textbf{0.2} & \textbf{0.3} & \textbf{0.4} & \textbf{0.5}  \\ \hline \hline
\multirow{9}{*}{\textbf{Porto}} & EDR & 57.173 & 203.993 & 806.033 & 2286.821 & 4872.231  \\
 & EDwP &  8.442 & 10.968 & 18.727 & 28.394 & 68.061 \\
 & Hausdorff & 10.026 & 23.293 & 56.561 & 89.827 & 275.206 \\
 & Fr\'echet & 10.668 & 18.516 & 29.740 & 93.851 & 181.271 \\ \cline{2-7} 
 & t2vec & 4.786 & 8.461 & 19.689 & 35.219 & 115.364 \\
 & TrjSR & 7.941 & 15.746 & 151.948 & 549.108 & 1341.883 \\
 & E2DTC & 5.100 &	9.385 &	21.845 & 39.402 & 124.320\\
 & \rred{CSTRM} & \rred{24.794} & \rred{47.137} & \rred{123.124} & \rred{257.540} & \rred{687.262} \\
 & TrajCL & \textbf{1.026} & \textbf{1.191} & \textbf{1.513} & \textbf{3.847} & \textbf{36.352} \\ \hline \hline
\multirow{9}{*}{\textbf{Chengdu}} & EDR & 398.870 & 455.498 & 446.041 & 594.004 & 820.628 \\
 & EDwP & 35.954 & 35.553 & 34.232 & 38.848 & 41.424  \\
 & Hausdorff & 117.908 & 121.474 & 116.300 & 125.098 & 152.359 \\
 & Fr\'echet & 87.673 & 89.005 & 86.420 & 91.204 & 104.220  \\ \cline{2-7} 
 & t2vec & 19.539 & 20.168 & 20.037 & 23.662 & 30.864 \\
 & TrjSR & 20.193 & 27.302 & 57.337 & 88.716 & 111.768  \\
 & E2DTC & 21.247 & 21.936 & 21.515 & 25.387 & 35.381\\
 & \rred{CSTRM} & \rred{30.165} & \rred{34.347} & \rred{41.677} & \rred{53.707} & \rred{75.486} \\
 & TrajCL & \textbf{1.219} & \textbf{1.465} & \textbf{1.261} & \textbf{6.611} & \textbf{15.523} \\ \hline  \hline
\multirow{9}{*}{\textbf{Xi'an}} & EDR & 279.835 & 285.550 & 340.820 & 367.227 & 516.571 \\
 & EDwP & 4.038 & 7.047 & 10.499 & 20.807 & 25.631 \\
 & Hausdorff & 64.390 & 122.651 & 124.112 & 127.969 & 184.158 \\
 & Fr\'echet & 66.813 & 86.647 & 144.120 & 160.499 & 196.099 \\ \cline{2-7} 
 & t2vec & 9.929 & 10.710 & 15.098 & 22.184 & 22.493 \\
 & TrjSR & 85.815 & 114.777 & 140.970 & 147.401 & 336.613 \\
 & E2DTC & 12.411 & 11.918 & 26.242 & 18.267 & 28.326 \\
 & \rred{CSTRM} & \rred{13.153} & \rred{16.056} & \rred{25.374} & \rred{34.194} & \rred{47.146} \\
 & TrajCL & \textbf{1.198} & \textbf{1.371} & \textbf{1.414} & \textbf{2.162} & \textbf{2.446} \\ \hline \hline
 
 \multirow{9}{*}{\textbf{\rred{Germany}}} & \rred{EDR} & \:\:\, \rred{1368.829} & \rred{1379.489} & \rred{1375.261} & \rred{1380.517} & \rred{1389.433} \\
 & \rred{EDwP} & \rred{2.173} & \rred{2.509} & \rred{2.176} & \rred{2.191} & \rred{2.209} \\
 & \rred{Hausdorff} & \rred{2.514} & \rred{2.742} & \rred{4.353} & \rred{4.448} & \rred{5.627} \\
 & \rred{Fr\'echet} & \rred{2.358} & \rred{2.492} & \rred{3.735} & \rred{3.824} & \rred{4.642} \\ \cline{2-7} 
 & \rred{t2vec} & \rred{4.453} & \rred{6.736} & \rred{9.087} & \rred{9.470} & \rred{9.775} \\
 & \rred{TrjSR} & \rred{24.539} & \rred{30.318} & \rred{55.002} & \rred{68.070} & \rred{111.175}  \\
 & \rred{E2DTC} & \rred{11.595} & \rred{13.478} & \rred{15.843} & \rred{18.532} & \rred{19.134} \\
 & \rred{CSTRM} & \rred{-} & \rred{-} & \rred{-} & \rred{-} & \rred{-} \\
 & \rred{TrajCL} & \rred{\textbf{1.048}} & \rred{\textbf{1.050}} & \rred{\textbf{1.059}} & \rred{\textbf{1.418}} & \rred{\textbf{2.045}} \\ 
\hlineB{3}
\end{tabular}
}
\end{table}

\begin{table}[ht]
\centering
\caption{Mean rank ($\blacktriangledown$)  vs. distortion rate}
\label{tab:newsimi_distort}
\resizebox{\columnwidth}{!}{%
\begin{tabular}{l|l|rrrrr}
\hlineB{3}
\textbf{Dataset} & \textbf{Method} & \textbf{0.1} & \textbf{0.2} & \textbf{0.3} & \textbf{0.4} & \textbf{0.5}  \\ \hline \hline
\multirow{9}{*}{\textbf{Porto}} & EDR & 28.243 & 28.498 & 27.899 & 28.070 & 28.932 \\
 & EDwP & 7.591 & 7.166 & 7.038 & 7.235 & 7.236 \\
 & Hausdorff & 6.549 & 6.737 & 6.706 & 6.592 & 6.739 \\
 & Fr\'echet & 8.689 & 8.854 & 8.755 & 8.636 & 9.083 \\ \cline{2-7} 
 & t2vec & 3.212 & 3.487 & 3.981 & 3.897 & 3.999 \\
 & TrjSR & 4.781 & 5.087 & 35.144 & 6.194 & 7.201 \\
 & E2DTC & 3.348 & 3.678 &4.210 & 4.129 & 4.222\\
 & \rred{CSTRM} & \rred{20.860} & \rred{20.081} & \rred{22.081} & \rred{24.688} & \rred{26.243} \\
 & TrajCL & \textbf{1.022} & \textbf{1.154} & \textbf{1.076} & \textbf{1.091} & \textbf{1.039} \\ \hline \hline
\multirow{9}{*}{\textbf{Chengdu}} & EDR & 381.155 & 374.169 & 368.976 & 364.785 & 360.5494 \\
 & EDwP & 34.392 & 33.151 & 33.817 & 29.300 & 29.193 \\
 & Hausdorff & 117.189 & 116.779 & 117.954 & 116.999 & 117.850\\
 & Fr\'echet & 86.658 & 86.949 & 88.367 & 88.238 & 88.253 \\ \cline{2-7} 
 & t2vec & 19.263 & 16.919 & 19.262 & 19.044 & 19.436 \\
 & TrjSR & 21.408 & 22.898 & 54.451 & 20.376 & 24.377 \\
 & E2DTC & 20.627 & 18.012 & 20.629 & 20.705 & 20.758\\
 & \rred{CSTRM} & \rred{25.026} & \rred{27.530} & \rred{28.691} & \rred{29.349} & \rred{26.786} \\
 & TrajCL & \textbf{1.168} & \textbf{1.260} & \textbf{1.810} & \textbf{2.508} & \textbf{1.115} \\ \hline  \hline
\multirow{9}{*}{\textbf{Xi'an}} & EDR & 275.205 & 270.394 & 266.143 & 263.054 & 259.541 \\
 & EDwP & 16.545 & 7.587 & 16.371 & 17.833 & 35.977 \\
 & Hausdorff & 184.629 & 183.114 & 188.238 & 186.298 & 179.990 \\
 & Fr\'echet & 195.383 & 195.244 & 195.385 & 197.348 & 196.140 \\ \cline{2-7} 
 & t2vec & 11.045 & 11.912 & 10.522 & 12.834 & 12.233 \\
 & TrjSR & 64.139 & 82.476 & 89.274 & 106.198 & 80.282\\
 & E2DTC & 13.490 & 14.768 & 13.227 & 16.621 & 16.498\\
 & \rred{CSTRM} & \rred{15.261} & \rred{15.063} & \rred{13.253} & \rred{16.865} & \rred{13.924} \\
 & TrajCL & \textbf{1.331} & \textbf{1.376} & \textbf{1.420} & \textbf{1.470} & \textbf{1.268} \\ \hline \hline
 
 \multirow{9}{*}{\textbf{\rred{Germany}}} & \rred{EDR} & \:\:\, \rred{1373.985} & \rred{1372.984} & \rred{1373.981} & \rred{1373.966} & \rred{1373.944} \\
 & \rred{EDwP} & \rred{2.488} & \rred{2.489} & \rred{2.492} & \rred{2.489} & \rred{2.489} \\
 & \rred{Hausdorff} & \rred{5.587} & \rred{5.576} & \rred{5.573} & \rred{5.566} & \rred{5.568} \\
 & \rred{Fr\'echet} & \rred{4.631} & \rred{4.625} & \rred{4.609} & \rred{4.625} & \rred{4.612} \\ \cline{2-7} 
 & \rred{t2vec} & \rred{3.863} & \rred{3.976} & \rred{4.903} & \rred{3.580} & \rred{3.625} \\
 & \rred{TrjSR} & \rred{27.146} & \rred{27.156} & \rred{27.032} & \rred{26.935} & \rred{27.035}  \\
 & \rred{E2DTC} & \rred{10.946} & \rred{11.161} & \rred{10.940} & \rred{11.275} & \rred{10.693} \\
 & \rred{CSTRM} & \rred{-} & \rred{-} & \rred{-} & \rred{-} & \rred{-} \\
 & \rred{TrajCL} & \rred{\textbf{1.049}} & \rred{\textbf{1.051}} & \rred{\textbf{1.049}} & \rred{\textbf{1.062}} & \rred{\textbf{1.054}} \\ 
\hlineB{3}
\end{tabular}
}
\end{table}

\emph{Varying distortion rate $\rho_{d}$.}
We further randomly distort the trajectories in $Q$ and $D$ by shifting point coordinates following Equation~\ref{eq:aug_pointshift}.
We vary the proportion of points distorted in each trajectory, denoted by $\rho_d$, from 0.1 to 0.5, and we keep $|D| = 100,000$.
Table~\ref{tab:newsimi_distort} shows the results, where \model\ outperforms all competitors again. Compared with the best baseline t2vec, \model\ reduces that mean rank of $T^Q_b$ by up to 2.85, 16.43, 27.37 and 3.67 times on the Porto, Chengdu, Xi'an and Germany datasets, respectively. The results further confirm that \model\ is more robust than the competitors on trajectories with distorted points.
We observe that the results of all methods may fluctuate when $\rho_d$ varies. 
This is because the random distortion is applied to all trajectories, not just the query or ground-truth ones. The relative similarity of the different trajectories may change towards any direction, such that there is no unified changing pattern of the mean rank of the target trajectory.

\begin{table}[ht]
\centering
\caption{\rred{Mean rank ($\blacktriangledown$) vs. test dataset}}
\label{tab:transfer}
\begin{tabular}{ll|r|r|r}
\hlineB{3}
\multicolumn{2}{l|}{} & \textbf{\rred{$\mathbf{|D|}$=100K}} & \textbf{\rred{$\mathbf{\rho_s}$=0.2}} & \textbf{\rred{$\mathbf{\rho_d}$=0.2}} \\ \hline \hline
\multicolumn{1}{l|}{\multirow{2}{*}{\textbf{\rred{Xi'an $\rightarrow$ Xi'an}} }} & \rred{t2vec} & \rred{8.644} & \rred{10.710} & \rred{11.912} \\
\multicolumn{1}{l|}{} & \rred{\model} & \textbf{\rred{1.107}} & \textbf{\rred{1.371}} & \textbf{\rred{1.376}} \\ \hline
\multicolumn{1}{l|}{\multirow{2}{*}{\textbf{\rred{Porto $\rightarrow$ Xi'an}} }} & \rred{t2vec} & \rred{1021.883} & \rred{1031.330} & \rred{6430.850} \\
\multicolumn{1}{l|}{} & \rred{\model} & \textbf{\rred{4.211}} & \textbf{\rred{8.295}} & \textbf{\rred{10.682}} \\ \hlineB{3}
\end{tabular}
\end{table}

\rred{\emph{Varying test dataset.}}
\rred{
We further study the generalizability of \model\ under a cross-dataset setting, i.e., training  \model\ on Porto and testing it on Xi'an without fine-tuning (denoted as Porto $\rightarrow$ Xi'an).  
We compare with the best learned competitor, t2vec, and we report the mean ranks of $T^q_b$ for the representative settings where 
$|D|$=100K (no trajectory modification), $\rho_s$=0.2 and $\rho_d$=0.2, respectively. 
We only present the results on Porto, since the relative performance on the other datasets is similar.
We also contrast the results with those on model training and testing both on Xi'an (denoted as Xi'an $\rightarrow$ Xi'an).}

\rred{As Table~\ref{tab:transfer} shows, 
\model\ consistently outperforms t2vec 
under the cross-dataset setting, and the performance gap is even larger than that under the same-dataset setting. Comparing with the results reported in Tables~\ref{tab:newsimi_dbsize},~\ref{tab:newsimi_downsampling} and~\ref{tab:newsimi_distort} on Xi'an, 
\model\ (Porto $\rightarrow$ Xi'an) still outperforms most of the heuristic methods which are dataset independent, except for EDwP with a small gap (4.211 vs. 3.606 at $|D|=100K$). These results show the strong  generalizability of \model, attributing to our  dual-feature encoder which can capture generic correlation patterns between similar trajectories that translate across datasets. In comparison, t2vec uses $k$NN to compute the distance between cells, which is more vulnerable to a changed data distribution across datasets.}

\subsection{Efficiency of Similarity Computation}\label{sec:exp_efficiency} 
\textbf{Setup.} We report the training and testing  times of the different methods to compute the similarity between 1,000 query trajectories from $Q$ against 100,000 data trajectories in $D$, i.e., $10^8$ trajectory similarity computations in total. The heuristic methods are run on a 32-core CPU. The learning-based methods are trained on GPU, and tested on a 32-core CPU and on GPU (to observe the best performance) separately. 

\begin{table}[ht]
\centering
\vspace{1mm}
\caption{Training time of learning-based measures (second)}
\label{tab:training_time}
\begin{tabular}{l|r|r|r|r}
\hlineB{3}
 & \textbf{Porto} & \textbf{Chengdu} & \textbf{Xi'an} & \textbf{Germany} \\ \hline \hline
t2vec & 5,992 & 6,599 & 6,638 & \rred{852} \\
TrjSR & 31,983 & 32,088 & 32,137 & \rred{9,604} \\
E2DTC & 7,998 & 8,893 & 9,759 & \rred{1,856} \\
\rred{CSTRM} & \textbf{\rred{2,956}} & \textbf{\rred{3,070}} & \textbf{\rred{3,650}} & \rred{-} \\
TrajCL & 3,611 & 3,572 & 4,182 & \textbf{\rred{524}} \\ 
\hlineB{3}
\end{tabular}
\vspace{1mm}
\end{table}

\begin{table}[ht]
\centering
\caption{Trajectory similarity computation times  (second)} 
\label{tab:time}
\begin{tabular}{cl|r|r|r|r}
\hlineB{3}
\multicolumn{1}{l}{} & \textbf{} & \textbf{Porto} & \textbf{Chengdu} & \textbf{Xi'an} & \textbf{Germany} \\ \hline \hline
\multicolumn{1}{c|}{\multirow{9}{*}{CPU only}} & EDR & 734 & 2,007 & 3,451 & \rred{4,609} \\
\multicolumn{1}{c|}{} & EDwP & 31,956 & 180,381 & 305,219 & \rred{341,904} \\
\multicolumn{1}{c|}{} & Hausdorff & 663 & 1,244 & 1,911 & \rred{3,568} \\
\multicolumn{1}{c|}{} & Fr\'echet & 1,047 & 2,706 & 2,964 & \rred{4,175} \\ \cline{2-6} 
\multicolumn{1}{c|}{} & t2vec & 55 & 73 & 70 & \rred{61}\\
\multicolumn{1}{c|}{} & TrjSR & 1,390 & 1,348 & 1,289 & \rred{1,338}\\
\multicolumn{1}{c|}{} & E2DTC & 55 & 73 & 70 & \rred{61}\\
\multicolumn{1}{c|}{} & \rred{CSTRM} & \rred{111} & \rred{149} & \rred{149}  & \rred{-} \\
\multicolumn{1}{c|}{} & TrajCL & 126 & 160 & 153 & \rred{164} \\ \hline
\multicolumn{1}{c|}{\multirow{5}{*}{GPU only}} & t2vec & 34 & 35 & 36 & \rred{37} \\
\multicolumn{1}{c|}{} & TrjSR & 228 & 235 & 237 & \rred{226}\\
\multicolumn{1}{c|}{} & E2DTC & 34 & 35 & 36 & \rred{37}\\
\multicolumn{1}{c|}{} & \rred{CSTRM} & \textbf{\rred{11}} & \textbf{\rred{15}} & \textbf{\rred{14}} & \rred{-} \\
\multicolumn{1}{c|}{} & TrajCL & 13 & 17 & 16 & \textbf{\rred{18}} \\ 
\hlineB{3}
\end{tabular}
\vspace{1mm}
\end{table}

\textbf{Results.} \emph{Training.} 
\rred{As Table~\ref{tab:training_time} shows, 
\model\ is only slightly slower than CSTRM on Porto, Chengdu and Xi'an but faster than the other models. CSTRM uses the vanilla multi-head self-attention, which can be regarded as a simplified version of our DualMSM module and hence is faster to train (but is also much worse in model accuracy as shown above). On Germany, \model\ is the fastest (CSTRM triggers an out-of-memory error as mentioned above), taking less than  9 minutes to train. Note that all methods are faster on Germany than on the other three datasets, as the Germany training set has 30,000 trajectories while the other three datasets each has  200,000 trajectories for training. The heuristic methods do not require training and hence no training times are reported for them. }

\emph{Similarity computation.}
As reported in Table~\ref{tab:time}, \model\ is also the fastest to complete the $10^8$ trajectory similarity computations, taking less than 20 seconds  (0.2$\mu$s per computation) when powered by GPU. It achieves up to a $10^4$-time speedup against the heuristic methods (on Xi'an against EDwP).  When running on CPU,  \model\ is still at least 4.26 times (on Porto against Hausdorff) and up to $10^3$ times (on Xi'an against EDwP) faster than the heuristic methods. 

We also note from the tables: 
(1) TrjSR is the slowest learning-based method in both training and testing. This is because it stacks 13 convolutional layers which is expensive to compute. 
(2) t2vec and E2DTC share the same testing times, as 
E2DTC only uses its backbone encoder during testing, which is essentially a t2vec encoder. 
(3) t2vec (and E2DTC) is faster than \model\ on CPU  but is slower on GPU for testing. This is because t2vec has $l$ recurrent matrix computation steps, while \model\ can compute in one single step which better suits GPU. 
(4) The running times of the heuristic methods vary largely across different datasets, while those of the learning-based methods are much less impacted. This is because the heuristic methods are heavily impacted by the trajectory length which varies across datasets, while the learning-based methods use the same embedding size across datasets.

\subsection{Training Scalability}\label{subsec:exp_training}
\rred{
Next, we study the scalability of \model\ in training. We report the mean ranks and  training times for the settings where $|D|$=100K, $\rho_s$=0.2 and $\rho_d$=0.2 on Porto like above.}

\begin{figure}[ht]
    \captionsetup[subfloat]{farskip=1pt,captionskip=0.5pt}
    \centering
    \subfloat[{\rred{Mean rank ($\blacktriangledown$) vs. \#epochs}}~\label{fig:exp:training_epoch}]{
        \includegraphics[width=0.23\textwidth]{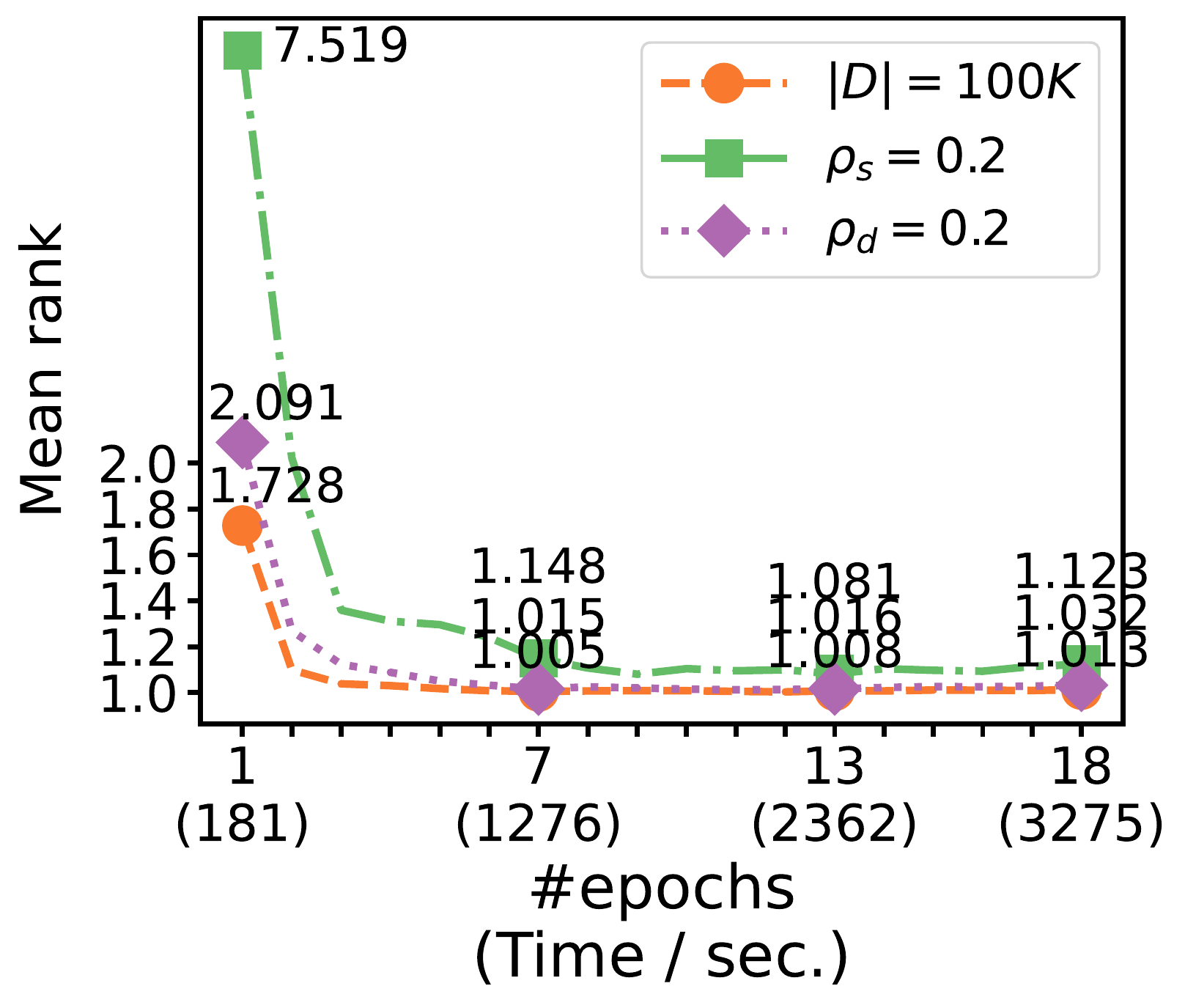}    
    }
    \subfloat[{\rred{Mean rank ($\blacktriangledown$) vs. \#trajectories }}~\label{fig:exp:training_numtrajs}]{
        \includegraphics[width=0.23\textwidth]{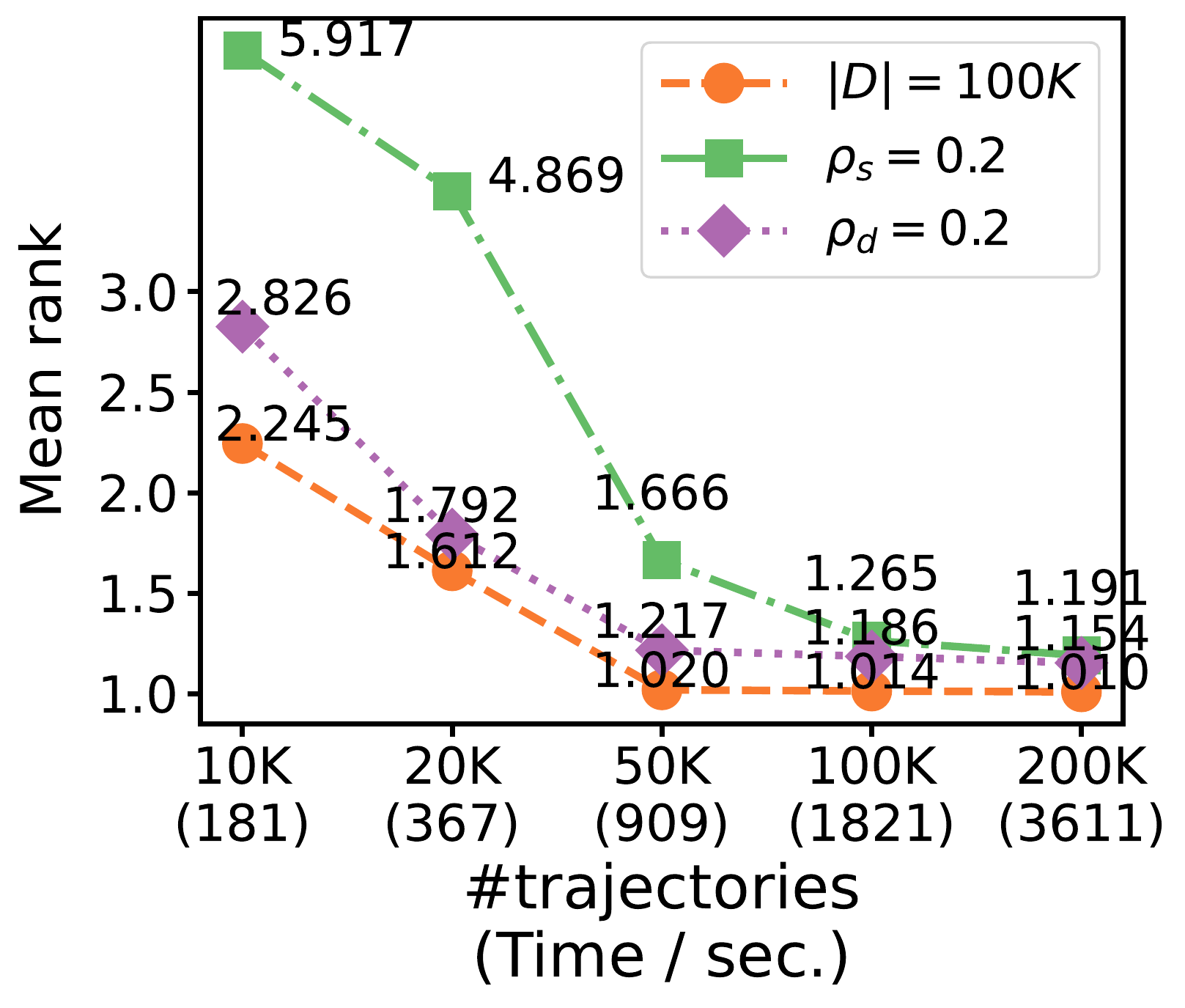}    
    }
    \caption{\rred{Impact of model training}}~\label{fig:exp:training}
    \vspace{-2mm}
\end{figure}

\rred{
\textbf{Impact of the number of the training epochs.}
Fig.~\ref{fig:exp:training_epoch} shows the learning curve of \model\ from 1 to 18 epochs of one run (when early termination is triggered and hence a time different from that in Table~\ref{tab:training_time} is reported).
As expected, \model\ produces lower mean ranks (i.e., better model accuracy) when it is trained for more epochs.  By the 7th epoch (about 20 minutes), \model\ has already achieved a satisfactory performance. This shows that \model\ is easy to train and converge, which helps its scalability.}

\rred{
\textbf{Impact of the number of training trajectories.} 
As Fig.~\ref{fig:exp:training_numtrajs} shows, \model\ benefits from more training trajectories. This is expected as more training trajectories offer more examples for the model to learn the different data patterns from. The performance gains diminish when using more than 50,000 original trajectories for training, while it takes a few more training trajectories when they are down-sampled or distorted, which is also intuitive. Using 50,000 trajectories, \model\ only takes about 15 minutes to train, which is quite practical. We used 200,000 as the default training set size, since the baseline methods require at least this size~\cite{t2vec,trjsr,e2dtc,cstrm}.}

\subsection{Efficiency of $K$ Nearest Neighbor Queries}\label{subsec:knn}
In real applications, we may index the trajectory database $D$ to support fast similarity searches.
We test \model\ under such a setting. To the best of our knowledge, this is the first reported results for $k$NN queries over trajectories using a non-trivial algorithm (i.e., non-full scans) based on  learned embeddings.

\textbf{Setup.}
We generate three trajectory databases $D$ with $|D| = 0.1, 1$ and $10$ million, by distorting ($\rho_{d} = 0.2$) randomly selected trajectories from the Xi'an dataset which has the largest number of points per trajectory. The three datasets have 11.8 million, 118.0 million and 1.2 billion trajectory points, respectively. We use the same 1,000 trajectory query sets as before, and run $k$NN queries over the generated databases. 

We run \model\ to generate embeddings for the data trajectories and index them with   
Faiss~\cite{faiss} which is a widely used library for similarity queries over dense vectors based on a Voronoi diagram. Note that our aim here is \emph{not} to come up with another trajectory index but to test the query efficiency of \model\ embeddings with existing $k$NN  algorithms.

We compare with Hausdorff, since \emph{the other learned methods will share the same query efficiency with \model\ on  Faiss}, while Hausdorff is the fastest heuristic measure (cf.~Table~\ref{tab:time}). 
For Hausdorff, we build a segment-based index with $k$NN pruning strategies, following a recent work DFT~\cite{DFT}.

\textbf{Results.} 
\emph{$k$NN query.} Fig.~\ref{fig:knn_querytime} shows the total response times to run 1,000 $k$NN queries, which grow with the dataset size $|D|$ for both methods, as expected. 
\model\ is about two orders of magnitude faster than Hausdorff, which attributes to both the fast embedding-based similarity computation and the efficient query procedure enabled by the embedding vectors. The Hausdorff index triggers an out-of-memory error when $|D| = 10$M and hence no results were obtained for this case. 

\begin{minipage}{\columnwidth}
    \vspace{2mm}
    \hspace{-3mm}
    \begin{minipage}[b]{0.44\columnwidth}
        \centering
        \includegraphics[width=\columnwidth]{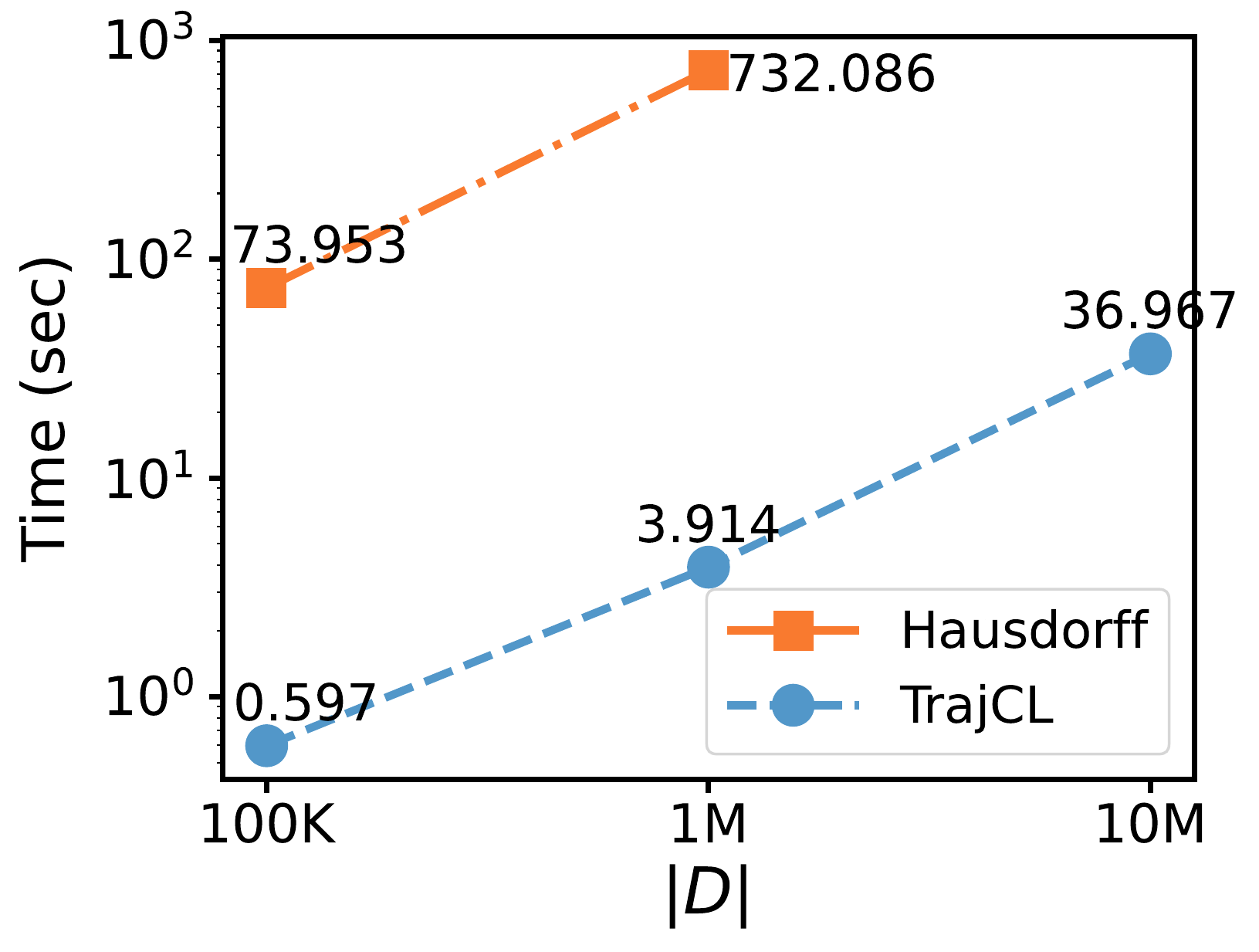}
        \captionof{figure}{$k$NN query costs}\label{fig:knn_querytime}
        \vspace{-15mm}
    \end{minipage}
    \begin{minipage}[b]{0.52\columnwidth}
        \centering
        \vspace{-15mm}
        \captionof{table}{Index building costs}\label{tab:knn_build}
        \resizebox{0.93\columnwidth}{!}{
        \begin{tabular}{l|r|rr}
        \hlineB{3}
         & \textbf{$|D|$} & \textbf{\begin{tabular}[c]{@{}c@{}}Time\\  (sec)\end{tabular}} & \textbf{\begin{tabular}[c]{@{}c@{}}RAM\\  (GB)\end{tabular}} \\ \hline \hline
        \multirow{3}{*}{Hausdorff} & 0.1M & 20.7 & 2.8 \\
         & 1M & 256.3 & 30.8 \\
         & 10M & - & OOM \\ \hline
        \multirow{3}{*}{TrajCL} & 0.1M & 42.7 & 0.5 \\
         & 1M & 426.1 & 2.9 \\
         & 10M & 4,234.0 & 20.3 \\ 
        \hlineB{3}
        \end{tabular}}
    \end{minipage}
    \vspace{2mm}
\end{minipage}

\emph{Index construction.} Table~\ref{tab:knn_build} further reports the index construction costs, which also grow with $|D|$. The \model\ index (i.e., Faiss) takes about twice the time of the  Hausdorff index (i.e., DFT) to build, where the extra times are spent on converting the trajectories to their embeddings. 
However,  the \model\ index takes much less memory than the Hausdorff index, e.g., 2.9 GB vs. 30.8 GB when $|D| = 1$M. This is because DFT needs to store auxiliary data for query pruning, 
which causes the out-of-memory error when $|D| = 10$M (1.2 billion segments, while DFT is a segment-based index).

\begin{table*}[ht]
\centering
\caption{HR@5, HR@20 and R5@20 ($\blacktriangle$) of self-supervised and supervised methods to approximate heuristic measures}
\label{tab:trajsimi}
\resizebox{\textwidth}{!}{%
\begin{tabular}{l|l|l|ccc|ccc|ccc|ccc|c}
\hlineB{3}
\multirow{2}{*}{\textbf{Dataset}} & \multirow{2}{*}{\textbf{Category}} & \multirow{2}{*}{\textbf{Method}} & \multicolumn{3}{c|}{\textbf{EDR}} & \multicolumn{3}{c|}{\textbf{EDwP}} & \multicolumn{3}{c|}{\textbf{Hausdorff}} & \multicolumn{3}{c|}{\textbf{Fr\'echet}} & \multirow{2}{*}{\begin{tabular}[c]{@{}c@{}}\textbf{Average} \\ \textbf{rank} ($\blacktriangledown$)\end{tabular}} \\ \cline{4-15} 
&  &  & \textbf{HR@5} & \textbf{HR@20} & \textbf{R5@20} & \textbf{HR@5} & \textbf{HR@20} & \textbf{R5@20} & \textbf{HR@5} & \textbf{HR@20} & \textbf{R5@20} & \textbf{HR@5} & \textbf{HR@20} & \textbf{R5@20} &  \\ \hline \hline
\multirow{9}{*}{\textbf{Porto}} & \multirow{6}{*}{\begin{tabular}[c]{@{}l@{}}Pre-trained \\ + fine-tuning\end{tabular}} & t2vec & 0.125 & 0.164 & 0.286 & 0.399 & 0.518 & 0.751 & 0.405 & 0.549 & 0.770 & 0.504 & 0.651 & 0.883 & 5\\
 &  & TrjSR & 0.137 & 0.147 & 0.273 & 0.271 & 0.346 & 0.535 & 0.541 & 0.638 & 0.880 & 0.271 & 0.356 & 0.523 & 8 \\
 &  & E2DTC & 0.122 & 0.157 & 0.272 & 0.390 & 0.514 & 0.742 & 0.391 & 0.537 & 0.753 & 0.498 & 0.648 & 0.879 & 6\\
 &  & \rred{CSTRM} & \rred{0.138} & \rred{0.191} & \rred{0.321} & \rred{0.415} & \rred{0.536} & \rred{0.753} & \rred{0.459} & \rred{0.584} & \rred{0.813} & \rred{0.421} & \rred{0.557} & \rred{0.768} & 3 \\
 &  & \model & 0.169 & 0.220 & 0.373 & 0.506 & 0.615 & 0.845 & 0.570 &  0.670 &  0.909 & 0.554 & 0.674 & 0.897 & 2\\ 
 &  & \model$^*$ & \textbf{0.172} & \textbf{0.222} & \textbf{0.376} & \textbf{0.546} & \textbf{0.646} & \textbf{0.881} & 0.643 & 0.721 & 0.954 & \textbf{0.618} & \textbf{0.740} & \textbf{0.955} & \textbf{1} \\ \cline{2-15} 
 & \multirow{3}{*}{Supervised} & Traj2SimVec & 0.119 & 0.163 & 0.285 & 0.172 & 0.253 & 0.390 & 0.339 & 0.429 & 0.543 & 0.529 & 0.664 & 0.894 & 9 \\
 &  & TrajGAT & 0.090 & 0.102 & 0.184 & 0.201 & 0.274 & 0.469 & \textbf{0.686} & \textbf{0.740} & \textbf{0.969} & 0.362 & 0.403 & 0.704 & 7 \\ 
 &  & T3S & 0.140 & 0.192 & 0.325 & 0.377 & 0.498 & 0.702 & 0.329 & 0.482 & 0.668 & 0.595 & 0.728 & 0.946 & 4 \\ \hline
\hline
\multirow{9}{*}{\textbf{Chengdu}} & \multirow{6}{*}{\begin{tabular}[c]{@{}l@{}}Pre-trained \\ + fine-tuning\end{tabular}} & t2vec & 0.144 & 0.250 & 0.332 & 0.285 & 0.364 & 0.539 & 0.347 & 0.510 & 0.681 & 0.479 & 0.607 & 0.823 & 6 \\
 &  & TrjSR & 0.137 & 0.245 & 0.347 & 0.210 & 0.251 & 0.430 & 0.555 & 0.690 & 0.887 & 0.369 & 0.472 & 0.671 & 7 \\
 &  & E2DTC & 0.136 & 0.241 & 0.317 & 0.269 & 0.348 & 0.518 & 0.319 & 0.484 & 0.647 & 0.461 & 0.585 & 0.802  & 8 \\
 &  & \rred{CSTRM} &  \rred{0.153} & \rred{0.239} & \rred{0.354} & \rred{0.342} & \rred{0.391} & \rred{0.642} & \rred{0.541} & \rred{0.665} & \rred{0.871} & \rred{0.519} & \rred{0.629} & \rred{0.848}  & 3 \\
 &  & \model & 0.171 & 0.258 & 0.382 & 0.366 & 0.405 & 0.665 &  0.594 &  0.715 &  0.918 & 0.602 & 0.697 & 0.918  & 2 \\ 
 &  & \model$^*$ & \textbf{0.178} & \textbf{0.271} & \textbf{0.404} & \textbf{0.381} & \textbf{0.429} & \textbf{0.671} & \textbf{0.697} & \textbf{0.785} & 0.968 & \textbf{0.726} & \textbf{0.794} & \textbf{0.981}  & \textbf{1} \\ \cline{2-15} 
 & \multirow{3}{*}{Supervised} & Traj2SimVec & 0.140 & 0.239 & 0.356 & 0.181 & 0.313 & 0.487 & 0.183 & 0.257 & 0.409 & 0.240 & 0.325 & 0.594  & 9 \\
 &  & TrajGAT & 0.091 & 0.188 & 0.261 & 0.257 & 0.373 & 0.612 & 0.695 & 0.771 & \textbf{0.970} & 0.435 & 0.496 & 0.853  & 5 \\ 
 &  & T3S & 0.149 & 0.237 & 0.346 & 0.330 & 0.417 & 0.605 & 0.440 & 0.582 & 0.741 & 0.498 & 0.622 & 0.831  & 4 \\ \hline
\hline
\multirow{9}{*}{\textbf{Xi'an}} & \multirow{6}{*}{\begin{tabular}[c]{@{}l@{}}Pre-trained \\ + fine-tuning\end{tabular}} & t2vec & 0.162 & 0.244 & 0.361 & 0.272 & 0.317 & 0.494 & 0.354 & 0.514 & 0.683 & 0.445 & 0.565 & 0.774  & 6 \\
&  & TrjSR & 0.151 & 0.267 & 0.391 & 0.218 & 0.273 & 0.439 & 0.536 & 0.661 & 0.843 & 0.379 & 0.464 & 0.685  & 7 \\
&  & E2DTC & 0.152 & 0.232 & 0.344 & 0.244 & 0.291 & 0.455 & 0.317 & 0.472 & 0.628 & 0.400 & 0.529 & 0.724  & 8 \\
&  & \rred{CSTRM} & \rred{0.161} & \rred{0.244} & \rred{0.360} & \rred{0.336} & \rred{0.364} & \rred{0.522} & \rred{0.522} & \rred{0.656} & \rred{0.848} & \rred{0.497} & \rred{0.605} & \rred{0.816} & 5 \\
&  & \model & 0.178 & 0.269 & 0.399 & 0.360 & 0.414 & 0.672 & 0.580 & 0.705 & 0.901 & 0.592 & 0.687 & 0.908  & 2 \\ 
&  & \model$^*$ & \textbf{0.181} & \textbf{0.277} & \textbf{0.413} & \textbf{0.362} & 0.424 & 0.677 & 0.695 & 0.779 & 0.964 & \textbf{0.690} & \textbf{0.769} & \textbf{0.966}  & \textbf{1}\\ \cline{2-15} 
& \multirow{3}{*}{Supervised} & Traj2SimVec & 0.143 & 0.255 & 0.388 & 0.163 & 0.287 & 0.491 & 0.130 & 0.217 & 0.372 & 0.156 & 0.254 & 0.487  & 9 \\
&  & TrajGAT & 0.131 & 0.269 & 0.387 & 0.312 & \textbf{0.440} & \textbf{0.696} & \textbf{0.739} & \textbf{0.787} & \textbf{0.976} & 0.476 & 0.537 & 0.884  & 3 \\
&  & T3S & 0.175 & 0.272 & 0.408 & 0.328 & 0.439 & 0.617 & 0.423 & 0.601 & 0.782 & 0.539 & 0.651 & 0.848  & 4 \\  \hline \hline

 \multirow{9}{*}{\textbf{\rred{Germany}}} & \multirow{6}{*}{\begin{tabular}[c]{@{}l@{}}\rred{Pre-trained} \\ \rred{+ fine-tuning}\end{tabular}} & \rred{t2vec} & \rred{0.050} & \rred{0.373} & \rred{0.382} & \rred{0.211} & \rred{0.260} & \rred{0.391} & \rred{0.202} & \rred{0.240} & \rred{0.357} & \rred{0.204} & \rred{0.257} & \rred{0.374} & \rred{7} \\
 &  & \rred{TrjSR} & \rred{0.029} & \rred{0.311} & \rred{0.346} & \rred{0.234} & \rred{0.288} & \rred{0.451} & \rred{0.445} & \rred{0.606} & \rred{0.780} & \rred{0.400} & \rred{0.573} & \rred{0.745} & \rred{6} \\
 &  &\rred{ E2DTC} & \rred{0.047} & \rred{0.338} & \rred{0.378} & \rred{0.198} & \rred{0.244} & \rred{0.369} & \rred{0.196} & \rred{0.235} & \rred{0.346} & \rred{0.200} & \rred{0.254} & \rred{0.369} & \rred{8} \\
 &  & \rred{CSTRM} & \rred{-} & \rred{-} & \rred{-} & \rred{-} & \rred{-} & \rred{-} & \rred{-} & \rred{-} & \rred{-} & \rred{-} & \rred{-} & \rred{-} &  \rred{9} \\
 &  & \rred{\model} & \rred{0.114} & \rred{0.406} & \rred{0.433} & \rred{0.444} & \rred{0.603} & \rred{0.740} & \rred{0.506} & \rred{0.612} & \rred{0.810} & \rred{0.531} & \rred{0.663} & \rred{0.857} & \rred{2} \\ 
 &  & \rred{\model$^*$} & \rred{\textbf{0.127}} & \rred{\textbf{0.427}} & \rred{\textbf{0.461}} & \rred{\textbf{0.486}} & \rred{\textbf{0.679}} & \rred{\textbf{0.908}} & \rred{\textbf{0.619}} & \rred{\textbf{0.736}} & \rred{\textbf{0.919}} & \rred{\textbf{0.620}} & \rred{\textbf{0.755}} & \rred{\textbf{0.922}} & \rred{\textbf{1}} \\ \cline{2-15} 
 & \multirow{3}{*}{\rred{Supervised}} & \rred{Traj2SimVec} & \rred{0.073} & \rred{0.386} & \rred{0.437} & \rred{0.309} & \rred{0.433} & \rred{0.584} & \rred{0.428} & \rred{0.634} & \rred{0.812} & \rred{0.456} & \rred{0.640} & \rred{0.883} & \rred{4} \\
 &  & \rred{TrajGAT} & \rred{0.081} & \rred{0.402} & \rred{0.442} & \rred{0.452} & \rred{0.648} & \rred{0.833} & \rred{0.563} & \rred{0.658} & \rred{0.889} & \rred{0.411} & \rred{0.537} & \rred{0.722} & \rred{3} \\ 
 &  & \rred{T3S} & \rred{0.044} & \rred{0.358} & \rred{0.365} & \rred{0.443} & \rred{0.590} & \rred{0.733} & \rred{0.423} & \rred{0.515} & \rred{0.657} & \rred{0.415} & \rred{0.564} & \rred{0.756} & \rred{5} \\

\hlineB{3}
\end{tabular}
}
\end{table*}

\subsection{Approximating Heuristic Measures}\label{subsec:exp_trajsimi}
We investigate the generalizability of \model\ by fine-tuning a pre-trained \model\ to learn and approximate a heuristic similarity measure with very few labelled data.
To the best of our knowledge, this is the first study to investigate the effectiveness of a learning-based trajectory similarity measure to approximate a heuristic measure. The fine-tuned \model\ can be used as a fast estimator for fast online computation of an expensive heuristic similarity measure (e.g., EDwP).

\textbf{Setup.}
We take the trained encoder of \model\ (and other self-supervised methods) on each dataset from Section~\ref{subsec:exp_newsimi} and connect it with a two-layer MLP where the size of each layer is the same as $d$. We fine-tune the last layer of the encoder and train the MLP to predict a given heuristic similarity value, optimizing the MSE loss. 
Besides the above self-supervised methods and the state-of-the-art supervised methods mentioned in Section~\ref{subset:exp_settings}, we also add an variant to show the optimal performance of \model\ where all layers of the encoder are fine-tuned, named \textbf{\model$^*$}.

Following the supervised  methods~\cite{traj2simvec,trajgat,t3s}, we report the hit ratio results \textbf{HR@$k$} ($k = 5, 20$), i.e., the ratio of the ground-truth top-$k$ trajectories in the predicted top-$k$ results, and  \textbf{R5@20}, which denotes the recall of returning the ground-truth top-5 trajectories in the predicted top-20 results.
We also report the \textbf{average rank} of each method over the 4 measures and 3 metrics on each dataset.

\textbf{Results.} 
Table~\ref{tab:trajsimi} summarizes the results. Overall, \model$^*$\ achieves the best similarity prediction accuracy (i.e., average rank is 1), while  \model\ ranks the second.

Comparing with the self-supervised baselines t2vec, TrjSR and E2DTC, \model$^*$\ and \model\ produce higher  HR@5, HR@20 and R5@20 scores consistently.
Based on HR@5, \model\ gains improvements by up to 128.0\%, 89.7\%, 41.5\% and 32.7\%  to approximate EDR, EDwP, Hausdorff and Fr\'echet, respectively. 
\model$^*$\ further improves over \model\ by up to 11.4\%, 9.4\%, 22.3\% and 20.6\% on approximating the four heuristic measures, respectively. 
Similar trends can be observed on HR@20 and R5@20.
The high R5@20 scores of \model$^*$\ and \model\ for Hausdorff and Fr\'echet, i.e., almost all over 0.9 on all four datasets, show that our models can approximate the two measures very well, since they are directly based on the spatial distances between point pairs.

Compared with the supervised methods Traj2SimVec, TrajGAT and T3S, \model$^*$\ is better in 80\% of the cases, i.e., 38 out of the 48 cases tested.
When \model$^*$\ performs better, its performance gain is 14.6\% on average. On the contrary, when a supervised baseline performs better, the average performance gap is just 2.8\%. Such cases are mostly observed on approximating Hausdorff, and the strong performance of TrajGAT in these cases is consistent with its own study~\cite{trajgat}.

These confirm that the pre-trained \model\ models can be easily adapted to approximate a given heuristic similarity measure. Such generalizability attributes to the trajectory augmentation methods and the \model\ encoder.

\subsection{Ablation Study}\label{subsec:ablation}
We study the impact of model components and augmentation methods in \model\ in this section. 

\textbf{Impact of the model components.}
We compare \model\ with two model variants: (1) \textbf{TrajCL-MSM} replaces DualMSM with the vanilla MSM used in Transformer. This variant also ignores the spatial features $\textbf{S}$. It can be regarded as applying a vanilla Transformer encoder in our proposed trajectory contrastive learning framework. (2) \textbf{TrajCL-concat} also uses the vanilla MSM, but it concatenates the spatial features  with the structural features, i.e., $\mathbf{T} \| \mathbf{S}$, as the input.  

We repeat the two sets of experiments of Sections~\ref{subsec:exp_newsimi} and~\ref{subsec:exp_trajsimi}, i.e., running our models on their own (denoted as \textbf{``no fine-tuning''}) and fine-tuning them to approximate a heuristic similarity measure (denoted as \textbf{``with fine-tuning''}). 
When running our models on their own, we follow the settings in Section~\ref{subsec:exp_training}. When fine-tuning our models to approximate a heuristic measure, we report HR@5. We only present the results on Porto for conciseness, since  similar comparative patterns are observed on the other datasets.

\begin{figure}[h]
    \captionsetup[subfloat]{farskip=1pt,captionskip=0.5pt}
    \centering
    \subfloat[{No fine-tuning ($\blacktriangledown$)}~\label{fig:exp:ablation_newsimi}]{
        \includegraphics[width=0.202\textwidth]{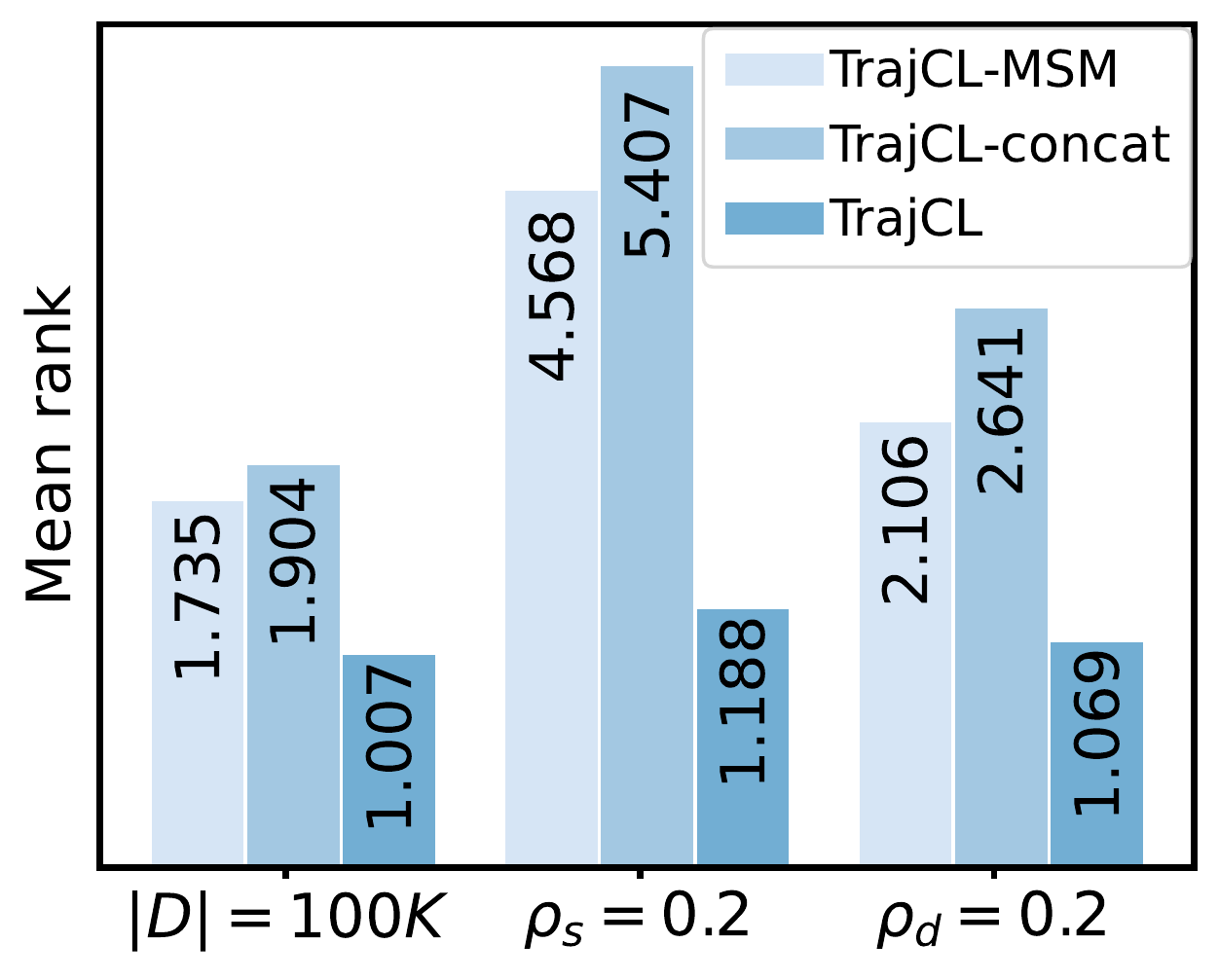}    
    }
    \subfloat[{With fine-tuning ($\blacktriangle$, HR@5)} ~\label{fig:exp:ablation_trajsimi}]{
    \includegraphics[width=0.24\textwidth]{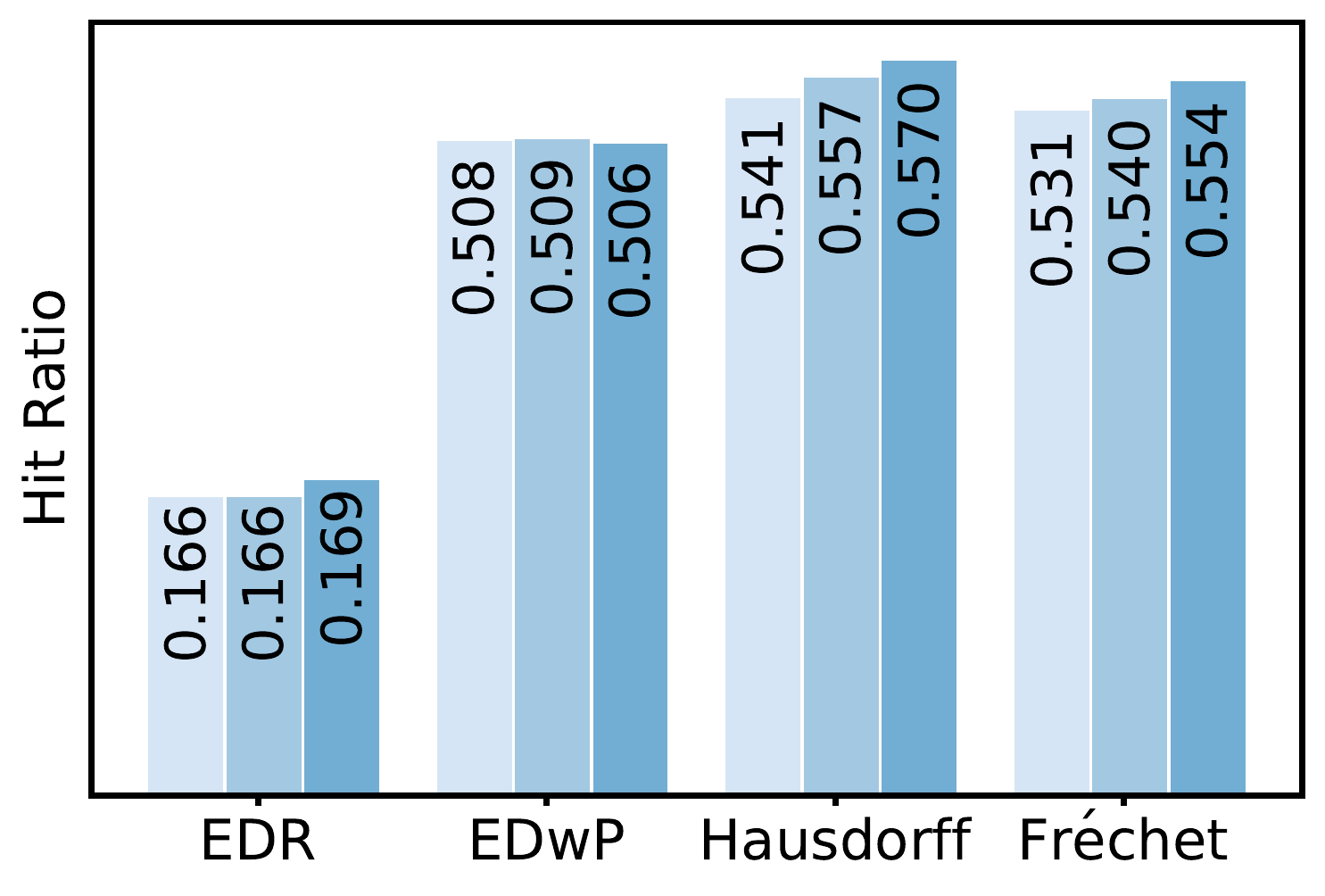}    
    }
    \caption{Ablation study results}~\label{fig:exp:ablation}
\end{figure}

\textbf{Results.} 
Fig.~\ref{fig:exp:ablation_newsimi} shows the results on \model\ variants without fine-tuning. 
\model\ performs better than the two variants by reducing the mean rank of $T^q_b$ by at least 72.29\% and 89.08\%, respectively. 
TrajCL-concat performs the worst, even though it uses the spatial features while TrajCL-MSM does not, as a direct concatenation can confuse the feature space of the model. The result highlights the importance of DualMSM that adaptively fuses both type of  input features.

Fig.~\ref{fig:exp:ablation_trajsimi} shows the results with fine-tuning.  \model\ still outperforms the two variants overall, except when approximating EDwP where all variants have similar results. This confirms that our DualMSM module has a strong generalization capability to capture the similarity between trajectories such that the fine-tuned \model\ can make more accurate predictions.

\textbf{Impact of the augmentation methods.} Next, we study how augmentation methods affect model performance by varying the augmentation methods to generate $\widetilde{T}$ and $\widetilde{T}'$. 
We use the same experimental setup as the last experiment. We only report the mean ranks on $|D|$=100K without fine-tuning and report the fine-tuning results of approximating EDwP (the most accurate but slowest heuristic measure), due to similar comparative patterns observed on other metrics.

\begin{figure}[ht]
    \captionsetup[subfloat]{farskip=1pt,captionskip=0.5pt}
    \centering
    
    \subfloat[{Mean rank ($\blacktriangledown$, $|D|$=100K)}~\label{fig:exp:augment_raw}]{
        \includegraphics[width=0.225\textwidth]{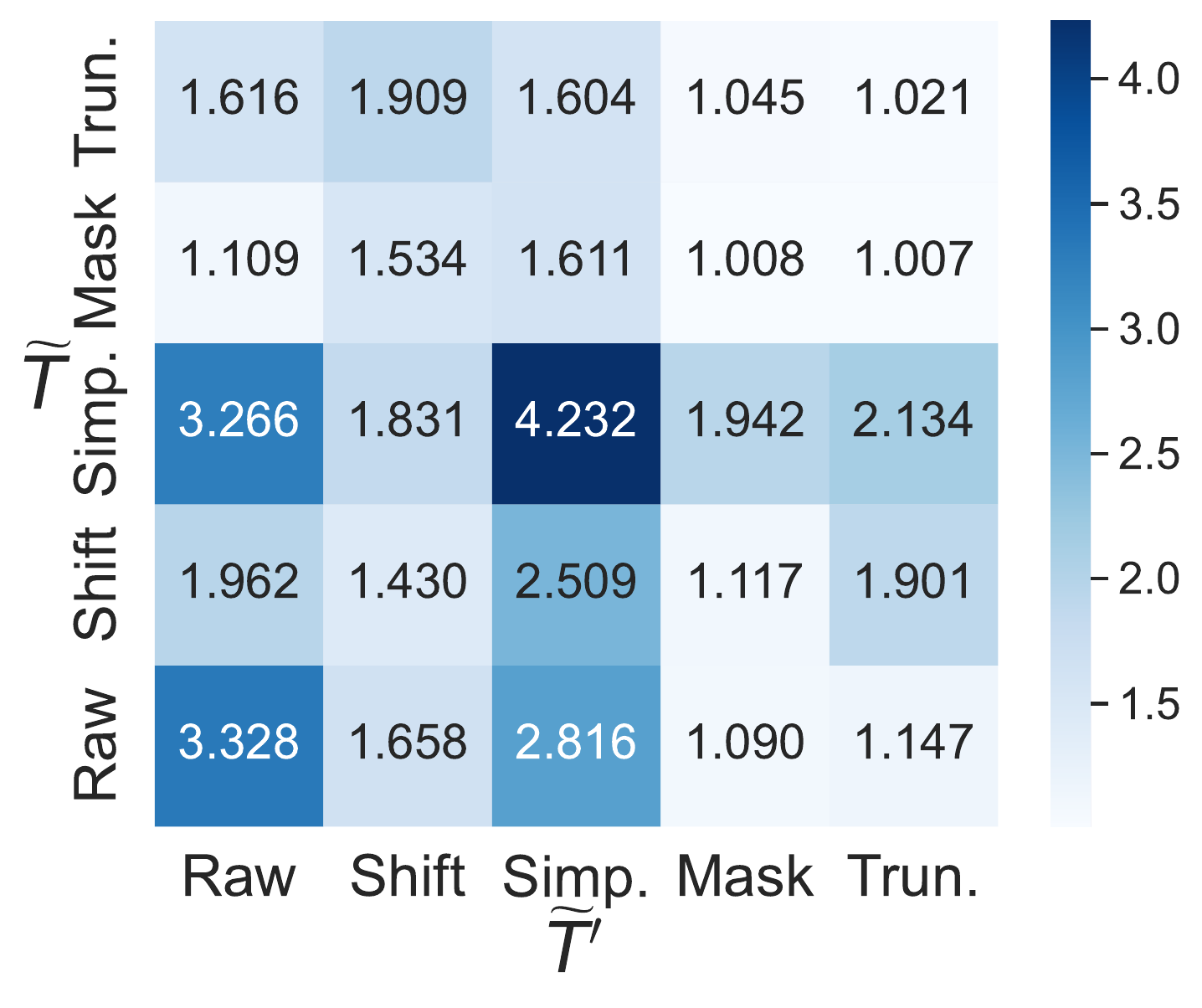}    
    }
    \subfloat[{HR@5 ($\blacktriangle$, with fine-tuning)}~\label{fig:exp:augment_hr5}]{
        \includegraphics[width=0.23\textwidth]{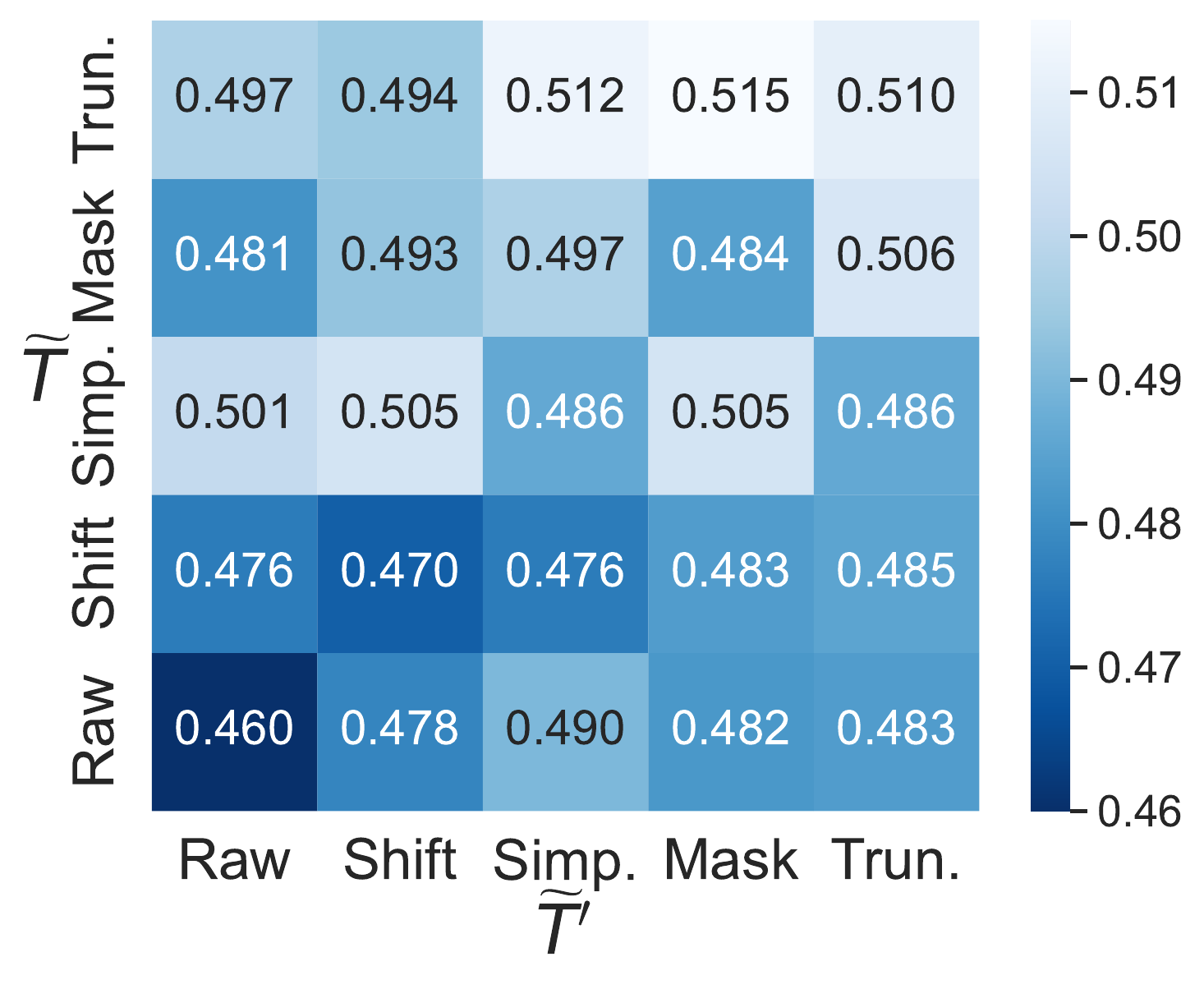}
    }
    \caption{Impact of the augmentation methods (Lighter color is better. \emph{Raw}, \emph{Shift}, \emph{Mask}, \emph{Trun.} and \emph{Simp.} denote no augmentation, point shifting, point masking, trajectory truncating  and trajectory simplification, respectively.)}\label{fig:exp:augment}
\end{figure}

\textbf{Results.} As Fig.~\ref{fig:exp:augment} shows, overall, augmentation helps \model\ learn more robust embeddings. 
\model\ without data augmentation (i.e., Raw\&Raw, using $T$ as $\widetilde{T}$ and $\widetilde{T}'$) has the lowest (i.e., worst) HR@5 value in Fig.~\ref{fig:exp:augment_hr5} and the second-largest (i.e., worst but second) mean rank values in Fig.~\ref{fig:exp:augment_raw}. Such results confirm the importance of the proposed augmentation methods.
Further, using the same augmentation methods for both $\widetilde{T}$ and $\widetilde{T}'$ may be sub-optimal, as this limits the learning space. 
Overall, point masking and trajectory truncating (i.e., Mask\&Trun.) produce the best similarity learning results, and thus have been used by default. 

\rred{Point masking helps learn the correlation between non-adjacent points and hence makes \model\ adaptive to trajectories with different sampling rates. Meanwhile, trajectory truncating helps learn the similarity between partial trajectories and the full ones.
In comparison, point shifting aims to guide \model\ to learn to overcome noises, while the grid cells used in \model\ can already achieve a similar purpose. 
Also, trajectory simplification  may remove too many points and hence miss key movement patterns to reflect trajectory similarity.}

\rred{
\textbf{Impact of parameters in  augmentation methods.}  
We focus on $\rho_d$ and $\rho_b$ for the two default augmentation methods point masking and trajectory truncating, respectively.}

\begin{figure}[ht]
    \captionsetup[subfloat]{farskip=1pt,captionskip=0.5pt}
    \centering
    
    \subfloat[{\rred{Mean rank ($\blacktriangledown$, $|D|$=100K)}}~\label{fig:exp:augment_masktrun_rate_raw}]{
        \includegraphics[width=0.225\textwidth]{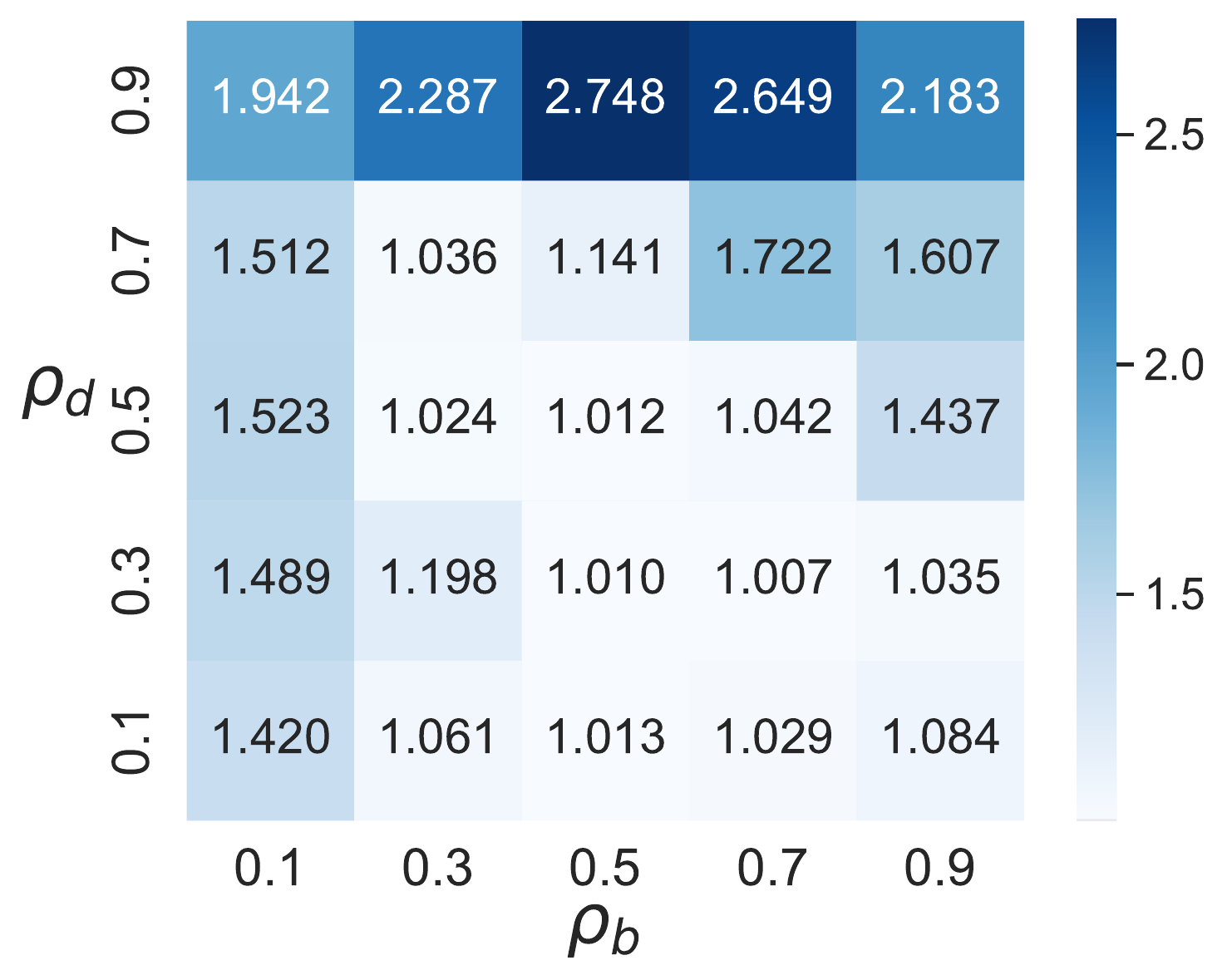}    
    }
    \subfloat[{\rred{HR@5 ($\blacktriangle$, with fine-tuning)}}~\label{fig:exp:augment_masktrun_rate_hr5}]{
        \includegraphics[width=0.23\textwidth]{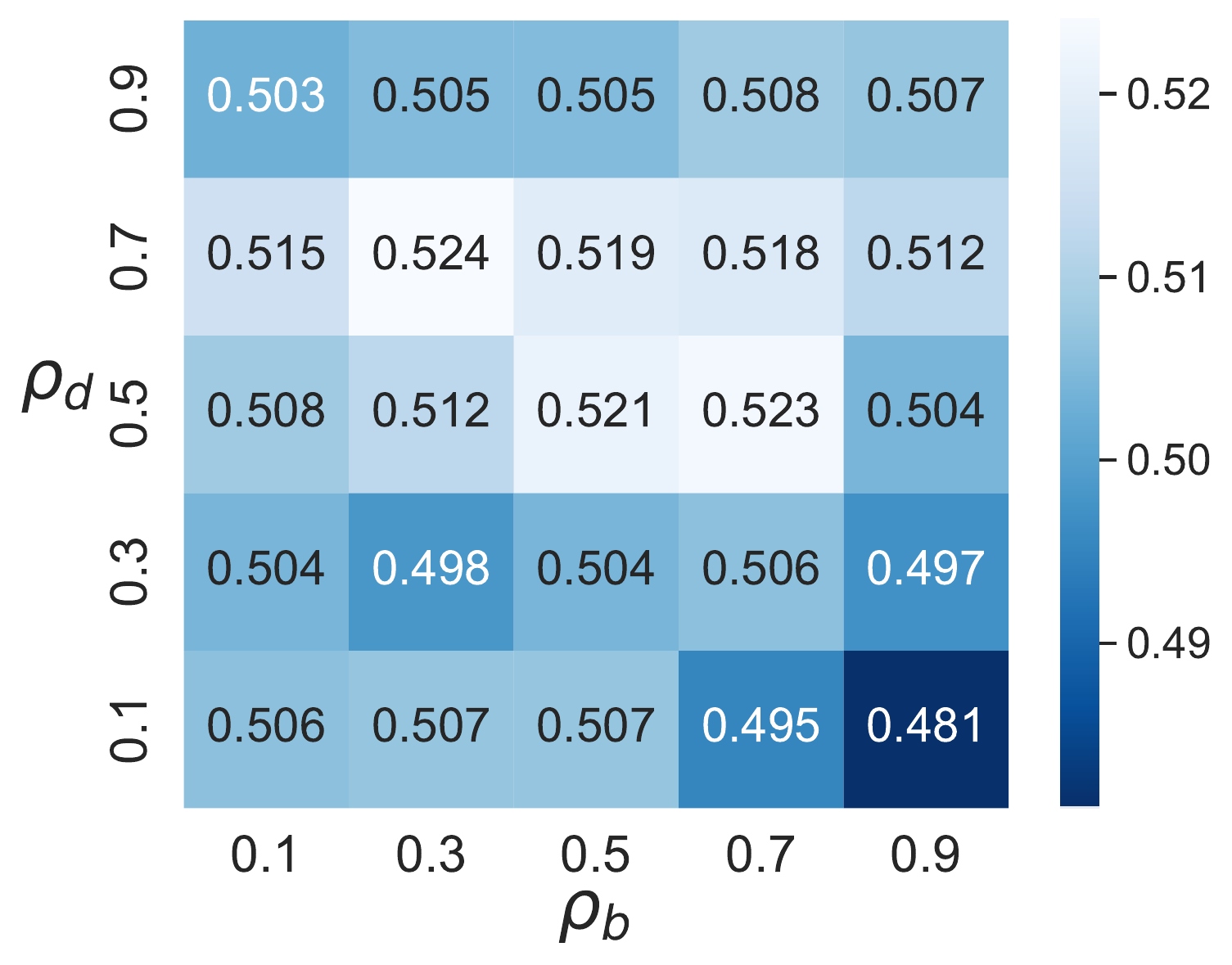}
    }
    \caption{\rred{Impact of parameters of the augmentation methods}}\label{fig:exp:augment_masktrun_rate}
\end{figure}

\rred{
\textbf{Results.}
We vary $\rho_d$ and $\rho_b$ from 0.1 to 0.9, respectively. As Fig.~\ref{fig:exp:augment_masktrun_rate} shows,
overall, \model\ is not heavily impacted by the two parameters unless for extreme values 0.1 and 0.9, i.e., when the augmented trajectories are too or little different from the original ones. When $\rho_d \in \{0.3,0.5\}$ and $\rho_b \in \{0.5,0.7\}$, \model\ performs the best. We use 0.3 and 0.7 by default for $\rho_d$ and $\rho_b$, respectively, where the lowest (i.e., best) mean rank is reached while HR@5 is also close to the best.
}

\subsection{Parameter Study}
We further study the impact of model parameters using  the same experimental setup as Section~\ref{subsec:ablation}.

\begin{figure}[ht]
    \captionsetup[subfloat]{farskip=1pt,captionskip=0.5pt}
    \centering
    \subfloat[{No fine-tuning ($\blacktriangledown$)}~\label{fig:exp:embedding_dim_newtraj}]{
        \includegraphics[width=0.23\textwidth]{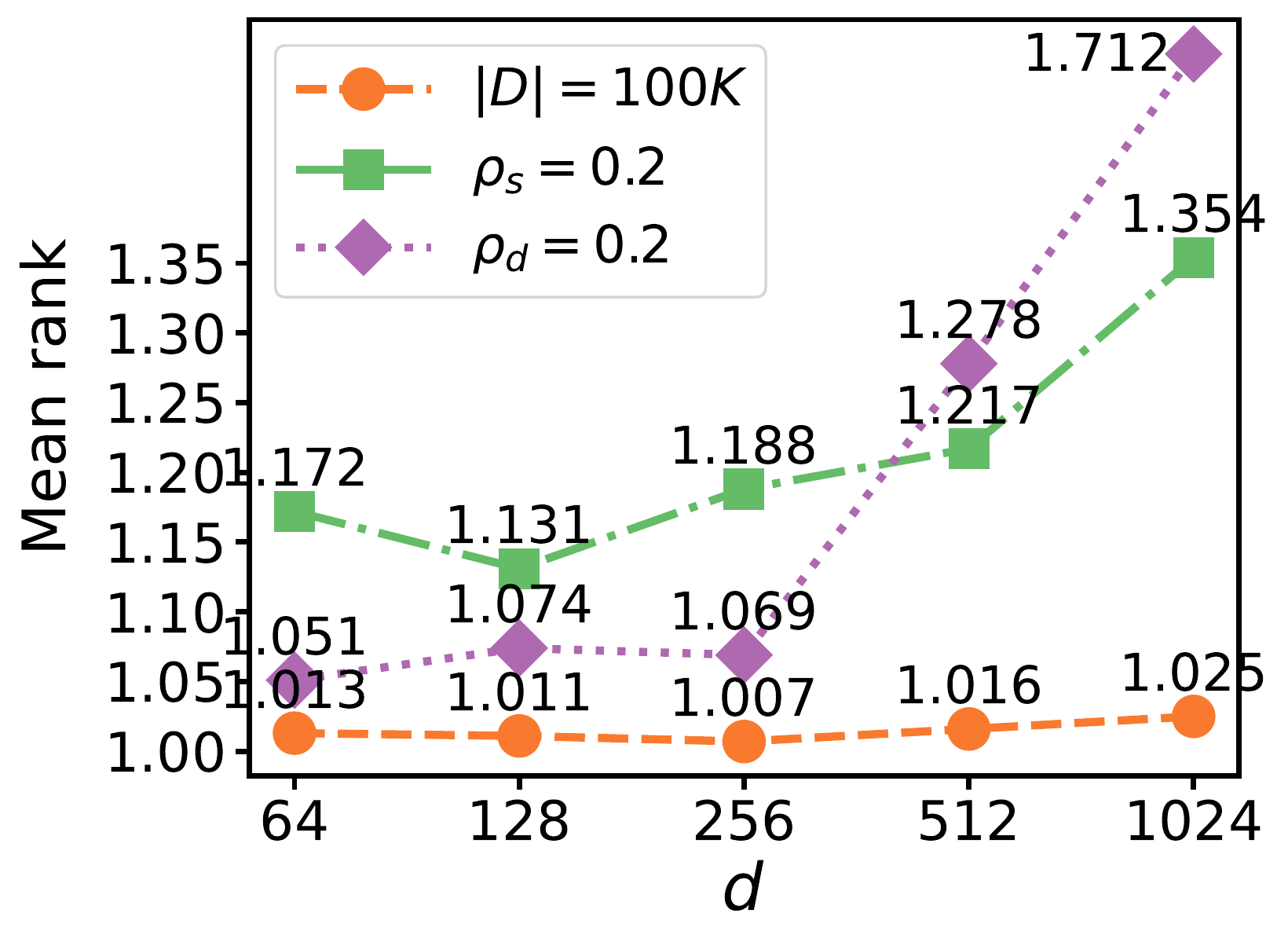}    
    }
    \subfloat[With fine-tuning ($\blacktriangle$)~\label{fig:exp:embedding_dim_trajsimi}]{
        \includegraphics[width=0.23\textwidth]{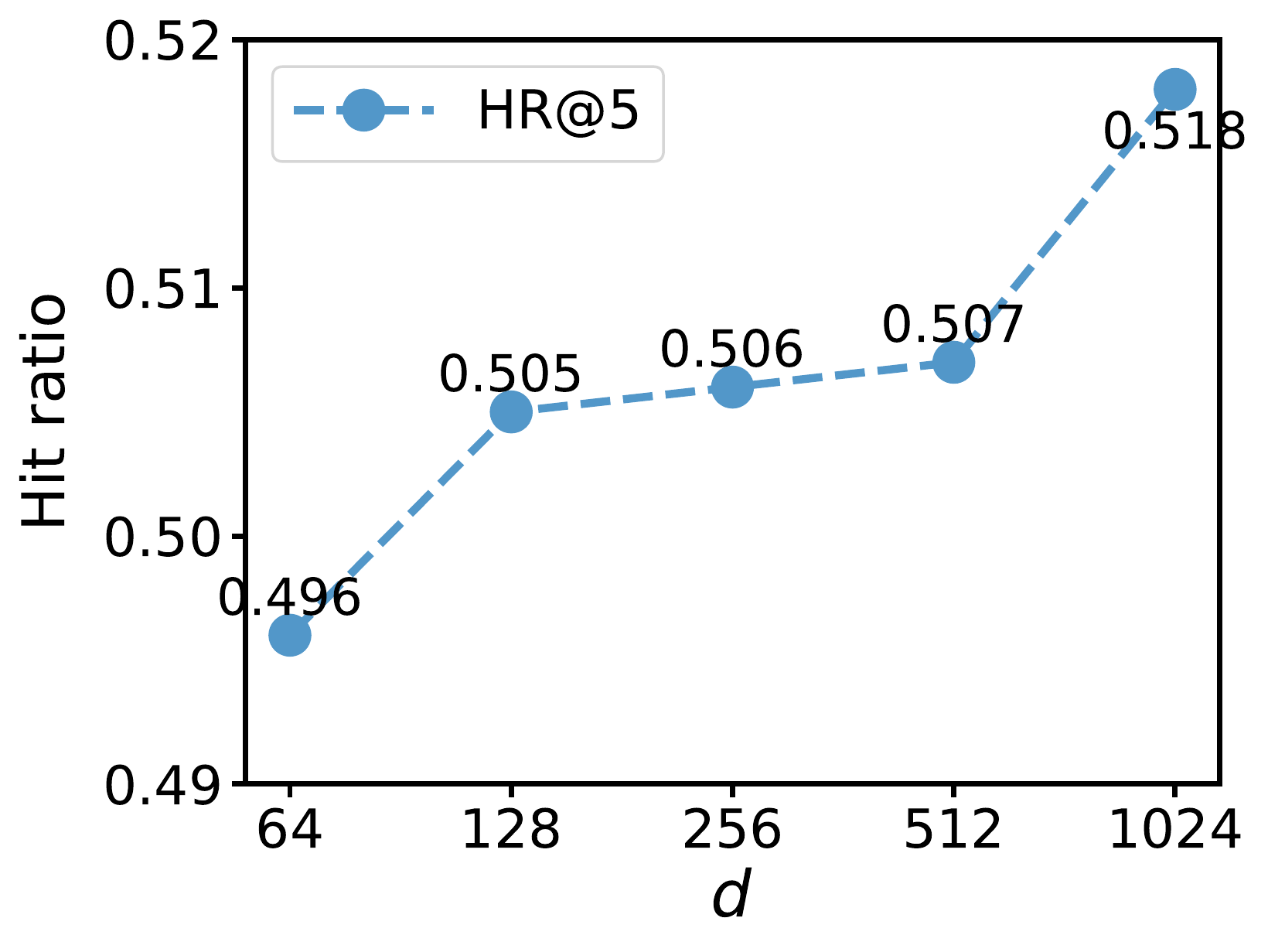}    
    }
    \caption{Impact of the embedding dimensionality}~\label{fig:exp:embedding_dim}
\end{figure}

\textbf{Impact of the embedding dimensionality $d$.} We vary $d$ from 64 to 1,024. 
As Fig.~\ref{fig:exp:embedding_dim} shows, 
overall, \model\ performs better when $d\leqslant256$ without fine-tuning. This is because a large embedding dimensionality results in  overfitting which impacts the similarity prediction results. 
When \model\ is fine-tuned, on the other hand, a larger $d$ allows it to better approximate heuristic measures. We note that a large $d$ also means larger fine-tuning and inference times. Thus, we have used 256 by default to balance model accuracy and efficiency.

\begin{figure}[th]
    \captionsetup[subfloat]{farskip=1pt,captionskip=0.5pt}
    \centering
    \subfloat[{No fine-tuning ($\blacktriangledown$)}~\label{fig:exp:layer_newsimi}]{
        \includegraphics[width=0.23\textwidth]{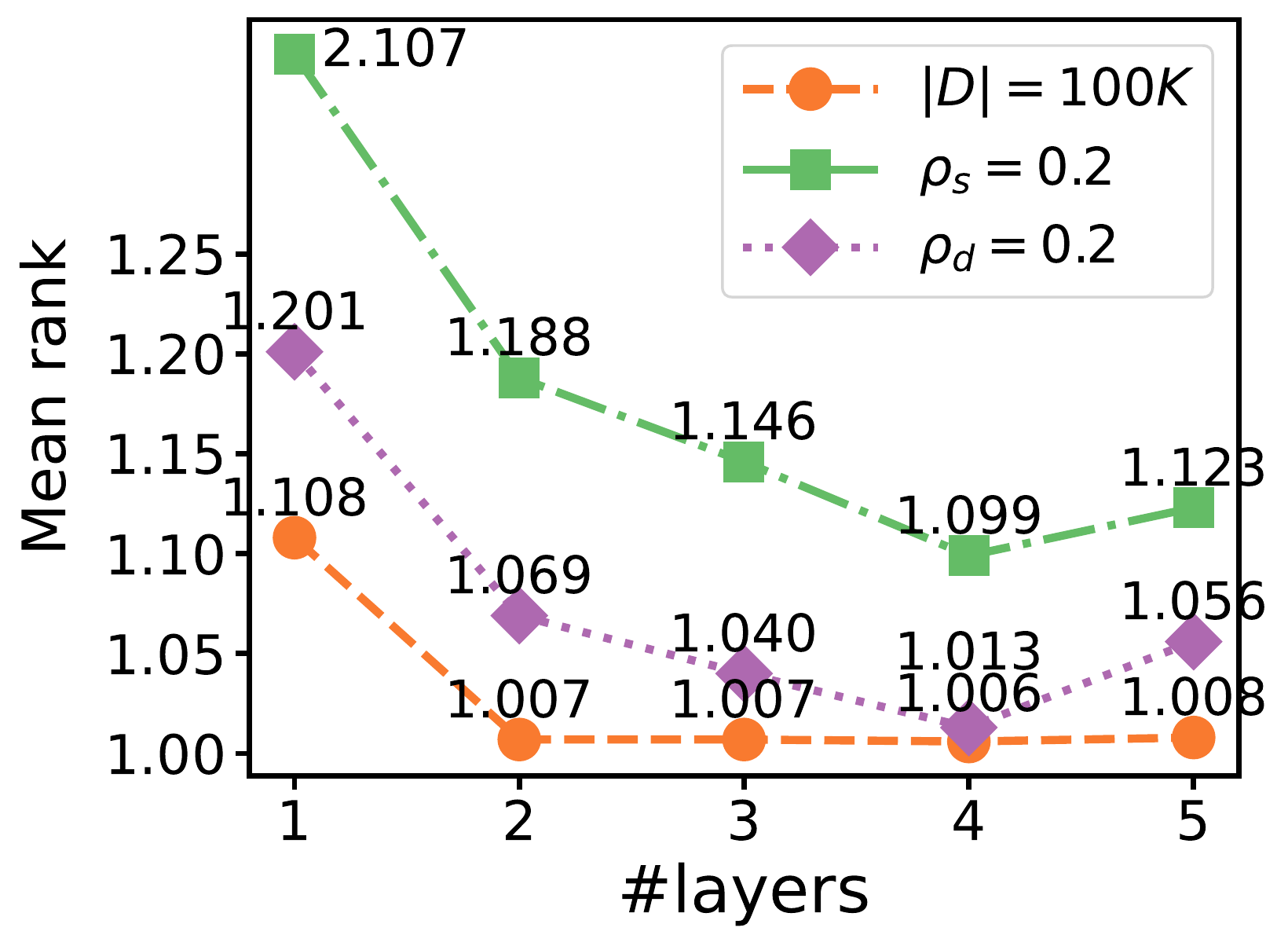}    
    }
    \subfloat[{With fine-tuning ($\blacktriangle$)} ~\label{fig:exp:layer_trajs}]{
    \includegraphics[width=0.23\textwidth]{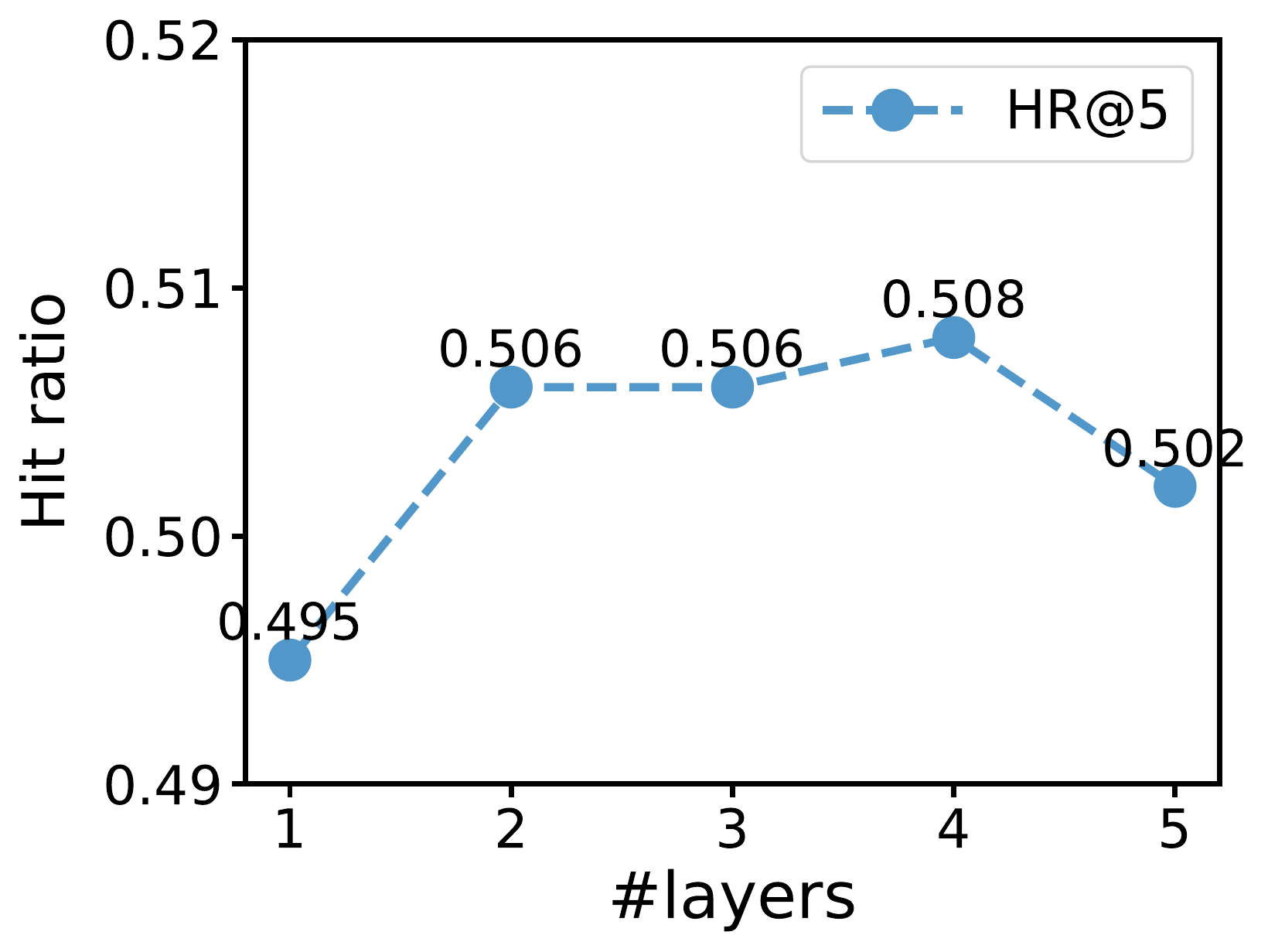}    
    }
    \caption{Impact of the number of the encoder layers}~\label{fig:exp:layer}
\end{figure}

\textbf{Impact of the number of the encoder layers $\#layers$.} Fig.~\ref{fig:exp:layer} shows the results for $\#layers \in [1, 5]$. 
As $\#layers$ increases, the model performance first improves and then drops when $\#layers$ exceeds 4. This is because adding more layers yields better feature generalization capability at first, since more layers help learn more complicated non-linear functions. When there are too many layers, however, it causes the model to overfit the training data. Given that a model with more layers takes substantially more time to train and run, we have only used a 2-layer encoder by default.

\begin{figure}[h]
    \captionsetup[subfloat]{farskip=1pt,captionskip=0.5pt}
    \centering
    \subfloat[{No fine-tuning ($\blacktriangledown$)}~\label{fig:exp:queue_newsimi}]{
        \includegraphics[width=0.23\textwidth]{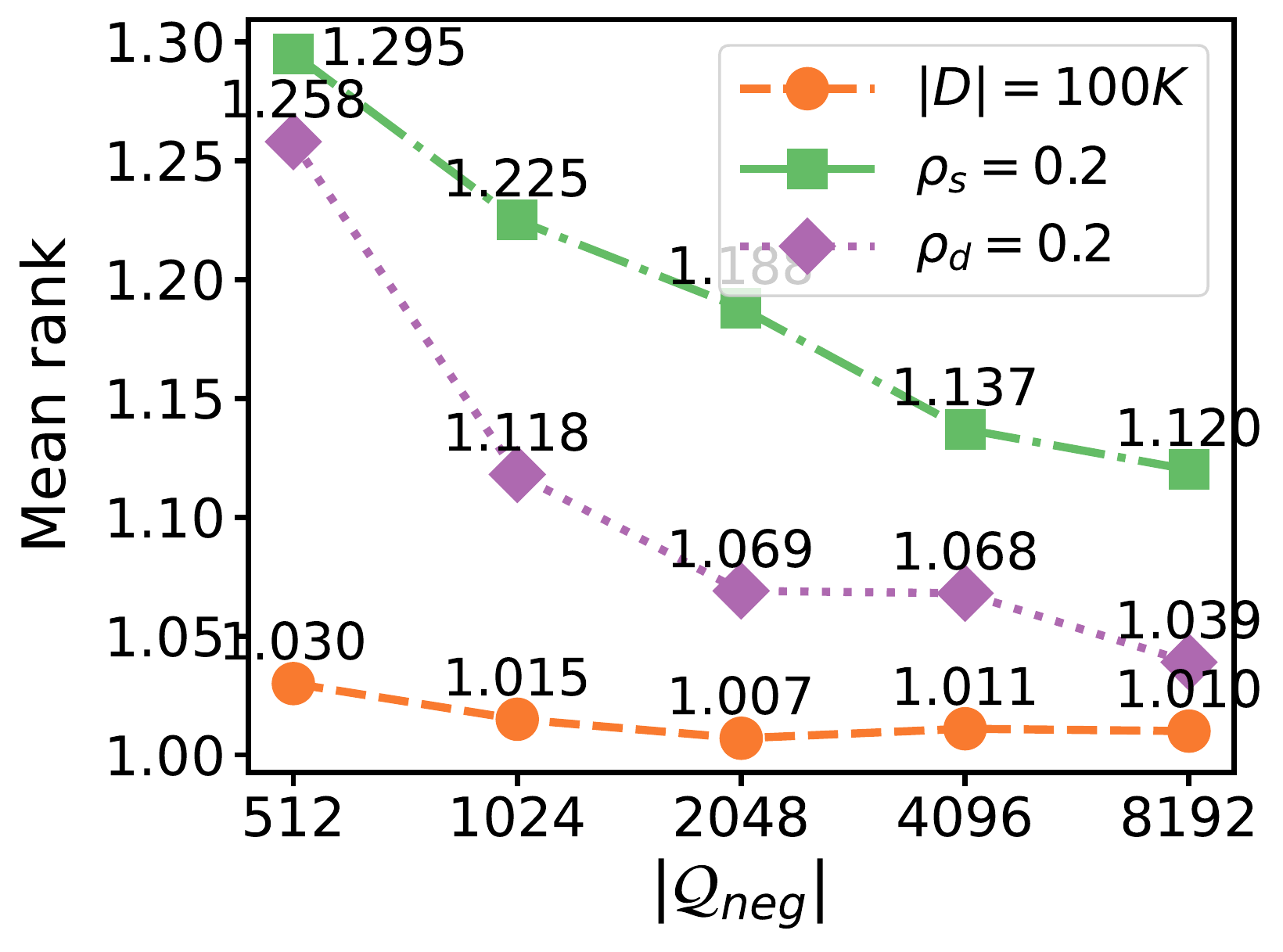}    
    }
    \subfloat[{With fine-tuning ($\blacktriangle$)}~\label{fig:exp:queue_trajsimi}]{
        \includegraphics[width=0.23\textwidth]{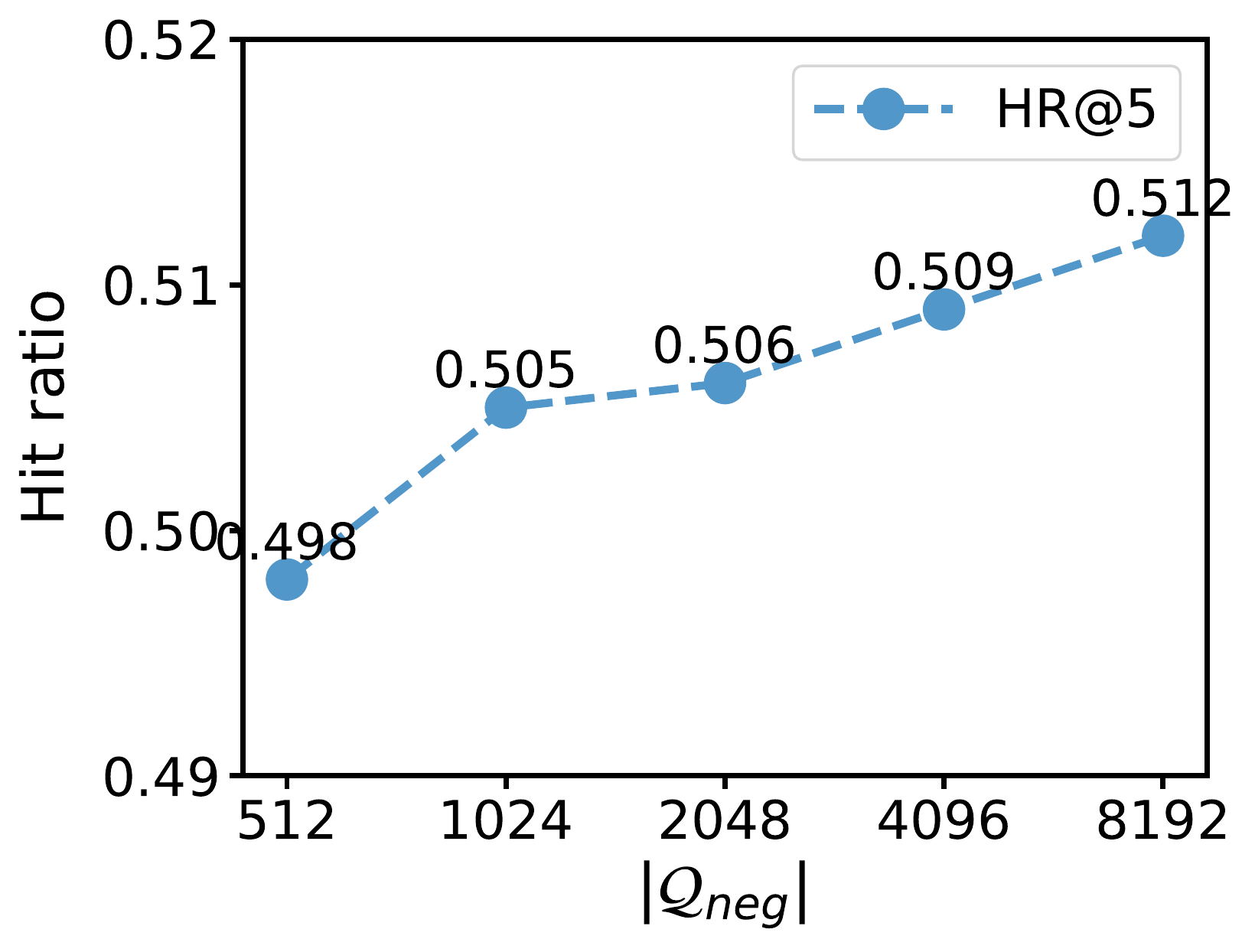}    
    }
    \caption{Impact of the size of the negative sample queue}~\label{fig:exp:queue}
\end{figure}

\textbf{Impact of the size of the negative sample queue $|Q_{neg}|$.} Lastly, we vary $|Q_{neg}|$ from 512 to 8,192. As Fig.~\ref{fig:exp:queue} shows, when $|Q_{neg}|$ increases, \model\  performs better in general, i.e., the mean ranks decrease and HR@5 increases. This is because more negative samples help reduce the bias caused by a small sample set, leading to more uniformly distributed embeddings in the latent space. 
As before, a large $|Q_{neg}|$ also leads to a larger training cost. We thus have used 2,048 by default.

\section{Conclusion}\label{sec:conclusion}
We proposed \model, a self-supervised trajectory similarity learning model that comes with a set of trajectory augmentation methods and a dual-feature multi-head self-attention-based trajectory backbone encoder. \model\ can learn the inherent similarity between trajectories and approximate predefined heuristic trajectory similarity measures. This makes \model\ a highly applicable trajectory similarity computation method.  
Experiments on four real trajectory datasets show that, compared with the state-of-the-art methods, \model\ achieves significant improvements in the accuracy for measuring trajectory similarity and approximating a heuristic measure.

\section*{Acknowledgment}
This work is partially supported by Australian Research Council (ARC) Discovery Project DP230101534. We thank Prof. Gao Cong for his comments which helped improve the paper.

\bibliographystyle{IEEEtran}
\bibliography{main}

\end{document}